\newif\ifanon
\ifanon
\documentclass[manuscript,screen,review,anonymous]{acmart}
\else
\documentclass[manuscript,screen,nonacm]{acmart}
\fi

\AtBeginDocument{%
  \providecommand\BibTeX{{%
    \normalfont B\kern-0.5em{\scshape i\kern-0.25em b}\kern-0.8em\TeX}}}

\setcopyright{rightsretained}
\copyrightyear{2024}
\acmYear{2024}
\acmDOI{}

\acmConference{}
\acmBooktitle{}
\acmISBN{}

\usepackage{subfig}
\usepackage{booktabs}

\usepackage{enumitem}
\setlist{leftmargin=7mm}

\begin{document}

   
\title{Exploring Keyboard Positioning and Ten-Finger Typing in Mixed Reality}

% \affiliation{%

\author{Cecilia Schmitz}
\affiliation{\institution{Michigan Technological University}
\city{Houghton}
\state{MI}
\country{USA}}
\email{cmschmit@mtu.edu}

\author{Joshua Reynolds}
\affiliation{\institution{Michigan Technological University}
\city{Houghton}
\state{MI}
\country{USA}}
\email{joshuare@mtu.edu}

\author{Scott Kuhl}
\affiliation{\institution{Michigan Technological University}
\city{Houghton}
\state{MI}
\country{USA}}
\email{kuhl@mtu.edu}

\author{Keith Vertanen}
\affiliation{\institution{Michigan Technological University}
\city{Houghton}
\state{MI}
\country{USA}}
\email{vertanen@mtu.edu}

%\renewcommand{\shortauthors}{Schmitz, Reynolds, Kuhl, and Vertanen}

% 150 words maximum
\begin{abstract}
Accuracy and speed are pivotal when typing. Mixed reality typing is typically performed by typing on a midair keyboard with your index fingers. This deprives users of both the tactile feedback available on physical devices and the ability to press keys with the most convenient finger. Our first experiment investigated providing tactile feedback by positioning the virtual keyboard on a table or wall. The keyboard was deterministic (without auto-correct), supported mixed case typing with symbols, and relied only on the hand-tracking provided by a commodity headset's egocentric cameras. Users preferred and had the highest entry rate of 12 words-per-minute using a midair keyboard. Error rates were similar in all conditions. Our second experiment explored ten-finger typing and used a novel eye-tracking technique to avoid accidental key presses. This technique was preferred for ten-finger typing and halved corrections. However, participants were faster using their index fingers without eye-tracking at 11 words-per-minute. 
\end{abstract}





\maketitle

\section{Introduction}
Mixed reality (MR) offers the ability to project virtual objects onto the real world and interact with them. As MR technology develops, users might wish to utilize MR to perform everyday tasks. Typing is critical to efficiently performing common tasks such as writing emails, browsing the web, and creating documents. Unlike a traditional keyboard where you can feel the locations of the keys and when they are actuated, MR systems often employ midair keyboards which lack tactile feedback. The loss of tactile feedback has been shown to impair users' accuracy when typing \cite{tabin2004tactile}. Further, MR systems typically restrict typing to just a user's index fingers. These issues can potentially reduce the utility of MR compared to conventional devices for supporting common, everyday computing activities. 

Speech input offers an alternative to typing but may not always be socially acceptable \cite{baier_acceptance_speech}, and recognition of difficult words  can be a problem \cite{adhikary2021text}. Auxiliary devices such as physical keyboards \cite{mcgill_dose,walker_physical, knierim2018physical, grubert_standard}, controllers \cite{speicher_selection, xu2019pointing}, gloves \cite{markussen2014vulture, mccaul2004predictive, he2022tapgazer}, or watches \cite{ahn_typing}, could help but require carrying additional devices. Other solutions re-imagine the text input process (e.g.~word-gesture keyboards \cite{kristensson_shark2, zhai_gesture, ahn_typing}, dwell-free gaze typing \cite{kristensson_dwellfree, hu2024dwellfree}, or finger gestures \cite{mccaul2004predictive, he2022tapgazer, feit2015investigating}). However, such methods introduce a learning curve that may be a barrier to MR adoption. Such methods also usually rely on a language model which restricts input to predictable natural language text. Compared to past work, we focus on mixed reality typing that: 1) leverages users' existing knowledge of tapping on an virtual keyboard, 2) uses only the sensors built into the headset, and 3) allows input of arbitrary character sequences (i.e.~without relying on a language model to compensate for noisy input).

Tactile feedback of a virtual keyboard on a table has been shown to improve performance~ \cite{dudley2019strategies}. What hasn't been shown is whether this is feasible using only the tracking available from the egocentric sensors on the headset and without relying on auto-correct. Further, only a keyboard positioned on a horizontal surface has been explored previously.

In Experiment 1, we explored three keyboard positions: midair, table aligned, and wall aligned. The table and wall surfaces add tactile feedback to keypresses and could be used whenever a flat surface is present. Although locating a keyboard on a surface may not be feasible in all situations, people are frequently in human-made environments where flat surfaces are prevalent. Participants typed using only their index fingers. They typed faster in midair at 12.2 words-per-minute (WPM) compared to on a table (9.1\,WPM) or a wall (8.8\,WPM). We found participants preferred the midair keyboard compared to keyboards near a table or wall.  

Experiment 2 focused on ten-finger typing, a feature requested by many participants in our first study. Pilot testing revealed ten-finger typing in midair led to numerous accidental keystrokes. Therefore, in addition to comparing ten- and index-finger typing, we also added a novel eye-tracking feature designed to prevent accidental keystrokes. Our feature filtered out key presses not located near a user's eye gaze location. We theorized since users were likely visually guiding their typing due to the lack of tactile feedback, this feature would not be an undue burden. We found users preferred index- to ten-finger typing. We found eye-tracking reduced accidental key presses, improved accuracy, and slightly increased entry rates for ten-finger typing (7.4\,WPM versus 7.1\,WPM).

Our contributions are as follows:\vspace{-1mm}
\begin{enumerate}
\item In two studies with 42 total participants, we found users could enter rich text (i.e.~mixed case with punctuation and numbers) accurately and at a reasonable speed of 11--12 words-per-minute on a deterministic midair keyboard.
\item While many users liked the tactile feedback afforded by a keyboard on a table or wall, we show midair typing provides higher performance on a current commercial headset due to challenges of hand-tracking near a surface.
\item Our novel eye-tracking based filter feature reduced unintentional key presses in midair ten-finger typing.
\item We release detailed logs from our studies and the source code of our HoloLens\footnote{HoloLens is an augmented reality headset released by Microsoft in 2016. A second version of the headset (and the one we used) was released in 2019.} experimental applications.
\end{enumerate}

\section{Related Work}
\subsection{Midair virtual keyboards}
For mixed reality to be useful for everyday computing tasks, text input needs to be efficient. Many  solutions have been proposed to address this need. A midair virtual keyboard allowing users to type directly with their hands is both familiar and easy-to-learn. Early systems such as ARKB \cite{lee_arkb} relied on external cameras to track a user's hands over the keyboard. Similar to our first study, ARKB used printed fiducial markers to position a keyboard horizontally and slightly above a table. No user study results were reported. \citet{markussen2013selection} also relied on external cameras for hand-tracking. Users either directly selected keys or used ambiguous gestures to specify keys. The keyboard was projected vertically in front of users on a large display. Direct typing was the fastest at 13\,WPM after six sessions. 

\citet{dudley_augmented} investigated midair typing on a HoloLens 1. The keyboard was located in midair slightly below a user's eyes and was tilted 20 degrees away from vertical. Similar to our work, they relied only on the built-in headset sensors. Despite the HoloLens 1 only tracking a user's wrist position, users typed at 18\,WPM on an auto-correcting keyboard with word predictions. In comparison to past work, we used a HoloLens 2 with full hand-tracking, eschewed the use of external cameras, and focused on the input of more difficult mixed case text with symbols and numbers.

\subsection{Physical keyboards in VR and AR}
Various work has explored typing on a physical keyboard while wearing a VR or AR headset. \citet{mcgill_dose} had users type: 1) without a headset, 2) wearing a headset with no view of the keyboard or a user's hands, 3) wearing a headset with a view of the keyboard and a user's hands blended in, or 4) wearing a headset with a view of the real-world. Users were fastest without the headset (59\,WPM) and slowest when they had no view of the keyboard or their hands (24\,WPM). Blending or full real-world visual feedback provided similar performance at 39 and 37\,WPM respectively. 

Rather than blending real-world video into VR, \citet{walker_physical} highlighted keys on a virtual keyboard as a user typed. Coupled with auto-correct, users typed in VR with only a virtual keyboard at 44\,WPM. \citet{chiossi_typing_mixed_reality} had users type in AR, VR, and augmented virtuality (where real-world imagery is displayed in VR \cite{mcgill_dose}). Users typed at similar speeds of 34.9, 32.5, and 34.5\,WPM respectively. While typing on a physical keyboard can be fast and support rich text input, users may not want to carry an extra device solely for text entry. As such, we focus on using a virtual keyboard.

\subsection{Tactile feedback}
While keyboard typing is a familiar interaction, the lack of tactile feedback on a midair keyboard can be a challenge. Tactile feedback significantly aids users even when typing on a physical keyboard. Typists with anesthetized fingertips were found to make seven times as many errors \cite{tabin2004tactile}. Tactile feedback can be added to these virtual keyboards via a handheld controller \cite{boletsis2019controllerbased, speicher2018selection}, a physical keyboard \cite{mcgill_dose,walker_physical, knierim2018physical, grubert_standard}, or devices placed on the fingers, hand, or wrist \cite{gupta2020investigating, pezent_tasbi, sakashita_wrist, gu2016dexmo}. But this requires users carry these devices in addition to the headset. Utilizing existing surfaces to provide tactile feedback avoids the needs for additional devices and has shown promising results. Users can type well on flat surfaces \cite{findlater2011flat} such as a tabletop display and on virtual keyboards projected onto a flat surface \cite{roeber2003canesta, dudley2019strategies}. 

\citet{dudley2019strategies} had users type with two or ten fingers on virtual keyboards located near a surface or in midair. Users' hands were tracked via an external camera array and markers on their hands. Users were seated at a desk and the midair keyboard was tilted slightly above horizontal. The system used an auto-correct algorithm based on the known target text. Users typed faster on the surface keyboards with either two fingers (56\,WPM versus 42\,WPM) or all ten fingers (52\,WPM versus 35\,WPM). Our work focuses on input without auto-correct and without knowing the target text.

\citet{richardson_decoding} also investigated surface typing by tracking users' hands via external cameras. Users typed phrases quickly at 73\,WPM. Using a neural network hand motion model and a language model, they typed sequences could be decoded at a low error rate of 2\%. In followup work, \citet{richardson_stegotype} used only the egocentric cameras of a VR headset to infer keypresses on a virtual keyboard situated on a desk. Users typed at 44\,WPM with an uncorrected error rate of 7\%. While the system had no explicit language model, the motion model may have learned one as the majority of its training data was English text. Indeed, on non-language text, offline experiments showed a much high error rate of 16\% compared to 5\% on English. Similar to \citet{richardson_stegotype}, we utilize the hand-tracking available from the egocentric cameras of a commodity headset. However, we rely solely on a human-in-the-loop process where the user observes their inferred finger locations and uses this feedback to supervise their typing. A system free from language-based prediction bias may be needed in some scenarios (e.g.~typing a password). 

\subsection{Interaction context}
Mixed reality offers the possibly of interaction in many real-world contexts, for example while sitting, standing, or walking. \citet{cheng_comfortable} explored if sitting at a table improved users' ability to select and drag targets in VR. Users' arms and finger were tracked via external cameras. They found having a table improved user accuracy and reduced fatigue of horizontal and vertical midair interfaces. \citet{dudley2019strategies} controlled for posture by having users sit at a desk while they typed on a virtual keyboard. We also chose to investigate seated interaction as this is a common context in which people do productivity related text entry. Further, given our chosen task of entering rich text without auto-correct, we wanted to reduce physical fatigue and user motion that could make hand-tracking interaction more challenging.

\subsection{Alternatives to QWERTY typing}
Moving somewhat away from the metaphor of tapping on a keyboard, a word-gesture keyboard \cite{kristensson_shark2,zhai_gesture} allows writing by swiping through a word's letters. The Vulture keyboard \cite{markussen2014vulture} allowed writing on a keyboard using a glove and an external camera array. After ten sessions, users wrote at 21\,WPM. Moving further away from conventional typing, an algorithm can infer a user's intended text from ambiguous gestures. Early work allowed users wearing a glove to write words in VR by selecting different sets of letters by flexing their finger \cite{mccaul2004predictive}. A similar recent system is TapGazer \cite{he2022tapgazer} which uses a glove to detect finger taps with disambiguation via eye gaze or additional taps.

Other approaches for text input in VR and AR have explored input methods other than hand gestures. The direction of a user's head can be used to select a letter from a circular menu \cite{xu_ringtext} or a QWERTY keyboard~\cite{yu2017techniques, speicher_selection, gizatdinova2018ultrasmall, xu_pointing}. While \citet{gizatdinova2018ultrasmall} found head-based typing was effective on small keyboards, users thought it felt unnatural and they preferred gaze-based typing. Gaze-based typing in VR has been investigated both as the sole input method \cite{xueshi2021handsfree} and in combination with head gestures \cite{wenxin2021hgaze}. However, the use of dwell time, the minimum time a user must look at a key in order to trigger it, can lead to eyestrain \cite{lystbaek2022freehand, pfeuffer2020menus}. An alternate approach, dwell-free eye typing \cite{kristensson_dwellfree}, allows users to gaze continually through letters. Using this approach, \citet{hu2024dwellfree} found users wrote at 12\,WPM using a HoloLens 2 headset. While potentially fast, dwell-free typing requires user learning and may not work for hard-to-predict text.

\subsection{Speech input}
Using speech input can help reduce MR challenges such as robustly sensing users hands and accurately displaying visual feedback. In early work, \citet{Bowman01092002} explored VR text input using Wizard of Oz speech recognition and input via a glove, keyboard, or tablet. Speech recognition was fastest at around 13\,WPM. In SWIFTER \cite{pick2016swifter}, users wrote in a virtual environment at 24\,WPM using speech with correction via a hand controller. Similar to our system, \citet{adhikary2021text} avoided auxiliary devices and relied solely on speech and hand-tracking for VR text input. Speaking and then correcting via hand gestures was the fastest at 28\,WPM. Notably, without speech, users wrote at 11\,WPM on a midair auto-correcting keyboard with word predictions. This entry rate was similar to our users' midair deterministic keyboard performance. While speech input can be a good option, it may not always be socially acceptable \cite{baier_acceptance_speech}, and accuracy can be poor in noisy environments or for challenging text (e.g.~usernames and passwords). 

\subsection{Ten finger interaction}
While existing MR systems often allow typing with both hands, typically they only support use of the index fingers \cite{lystbaek2022freehand, dudley_augmented, adhikary2021text}. Systems that allow use of all fingers generally require an auxiliary input device \cite{he2022tapgazer}, external sensing device \cite{yi_atk}, or physical keyboard \cite{mcgill_dose, walker_physical, knierim2018physical, grubert_standard}. Ten-finger typing is challenging as the dexterity limitations of the human hand can make some multi-finger gestures difficult without the tactile assistance of a physical keyboard \cite{sridhar2015dexterity}. Additionally, ten-finger typing on virtual keyboards can lead to accidental key coactivations \cite{foy2021coactivations}. While multi-finger gestures have been used for navigating MR interfaces, the small movements required for typing may be difficult for vision-based hand-tracking systems to recognize \cite{malik2005gestural}. Past work has showed that users naturally look at the keys they are pressing \cite{lystbaek2022freehand}. Thus in Experiment 2, we explore using eye-tracking to filter out key presses not near a user's gaze. 

\section{Experiment 1}

In our first experiment, participants wore a HoloLens 2 headset and typed sentences on a virtual keyboard using their index fingers. We measured user performance and preference for typing on a midair keyboard and keyboards located on a table and on a wall. Our primary goal was to assess the impact of different keyboard positions, including placement on a horizontal and vertical surface. We opted not to include a physical keyboard condition since typing on physical keyboards has been well-studied and our focus is on on-the-go applications where all sensing is performed by the headset. Since midair typing is currently a common choice for MR typing applications, we used a midair keyboard as a baseline.

\subsection{Design}

\begin{figure}[tb]
\centering
\includegraphics[height=3.5cm]{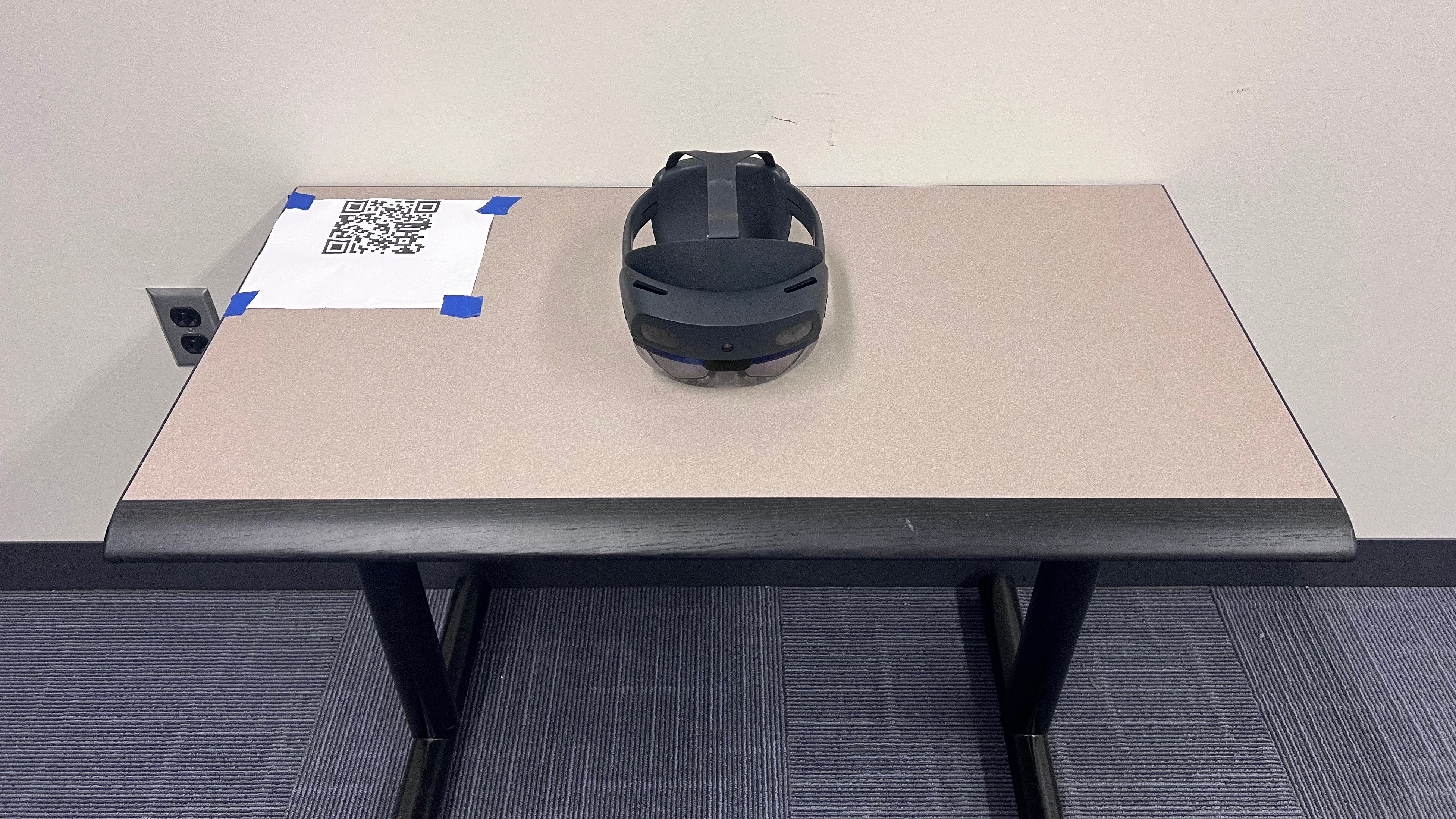}\hspace{2mm}
\includegraphics[height=3.5cm]{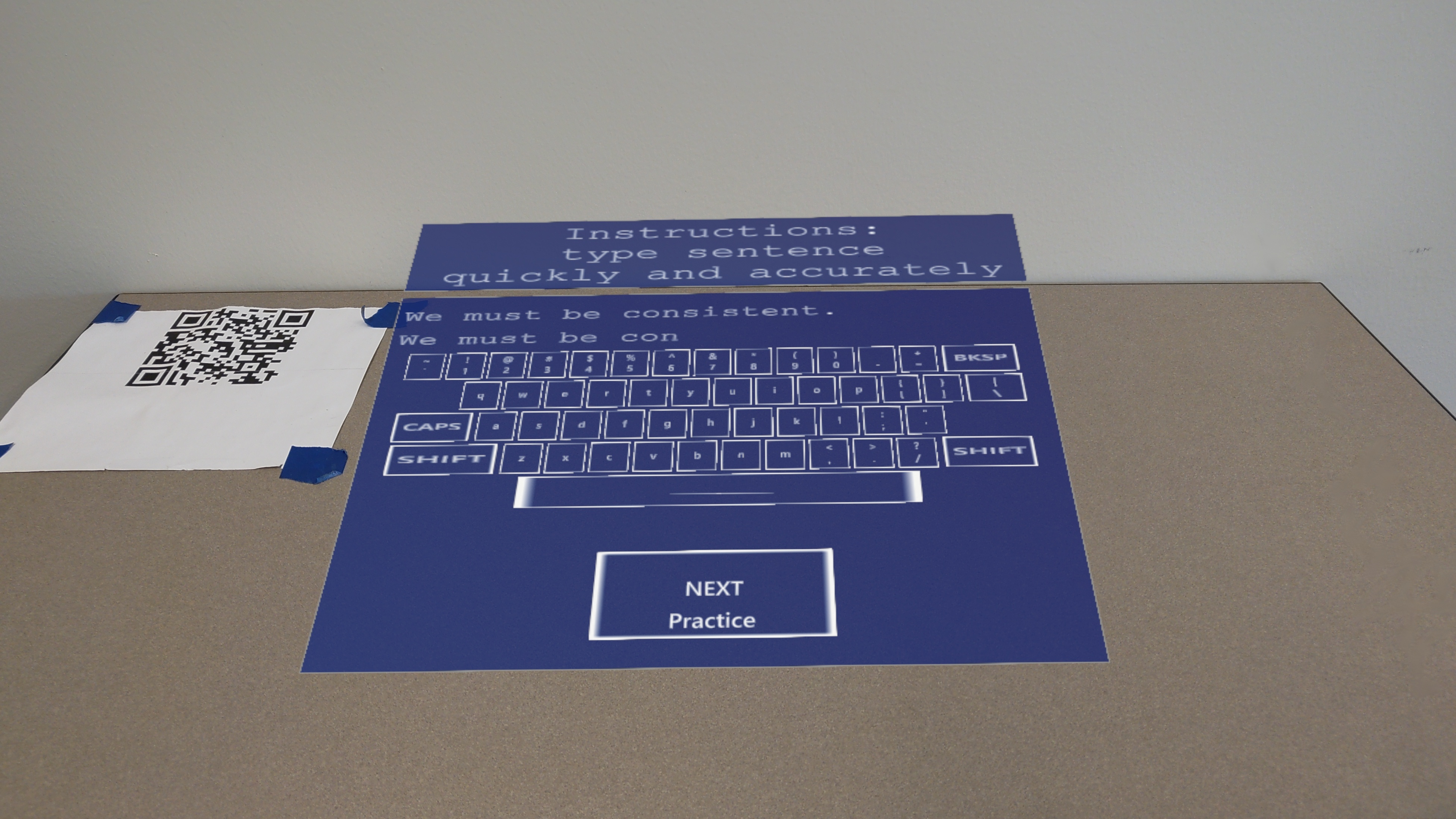}\\ \vspace{2mm}
\includegraphics[height=3.5cm]{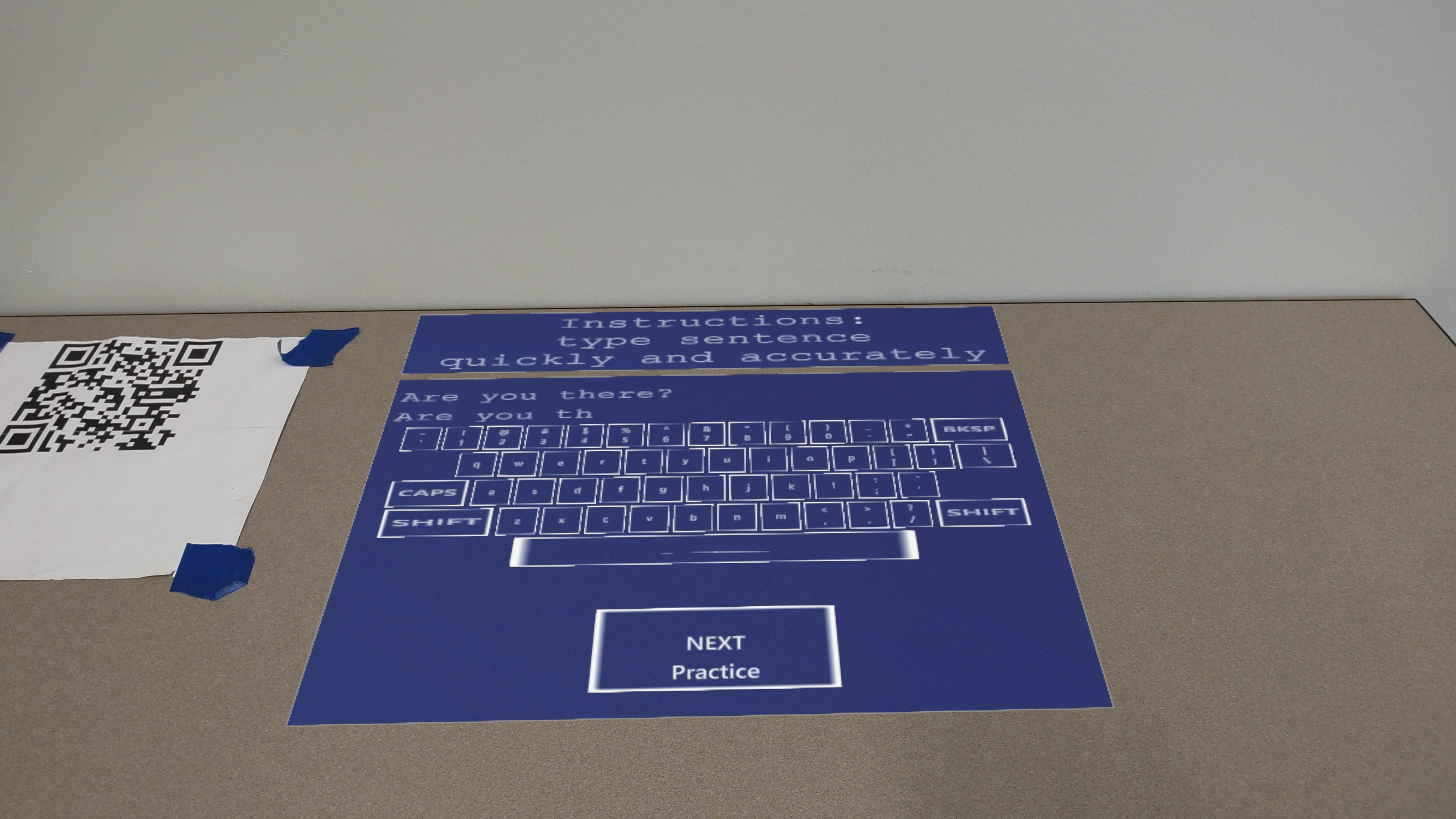}\hspace{2mm}
\includegraphics[height=3.5cm]{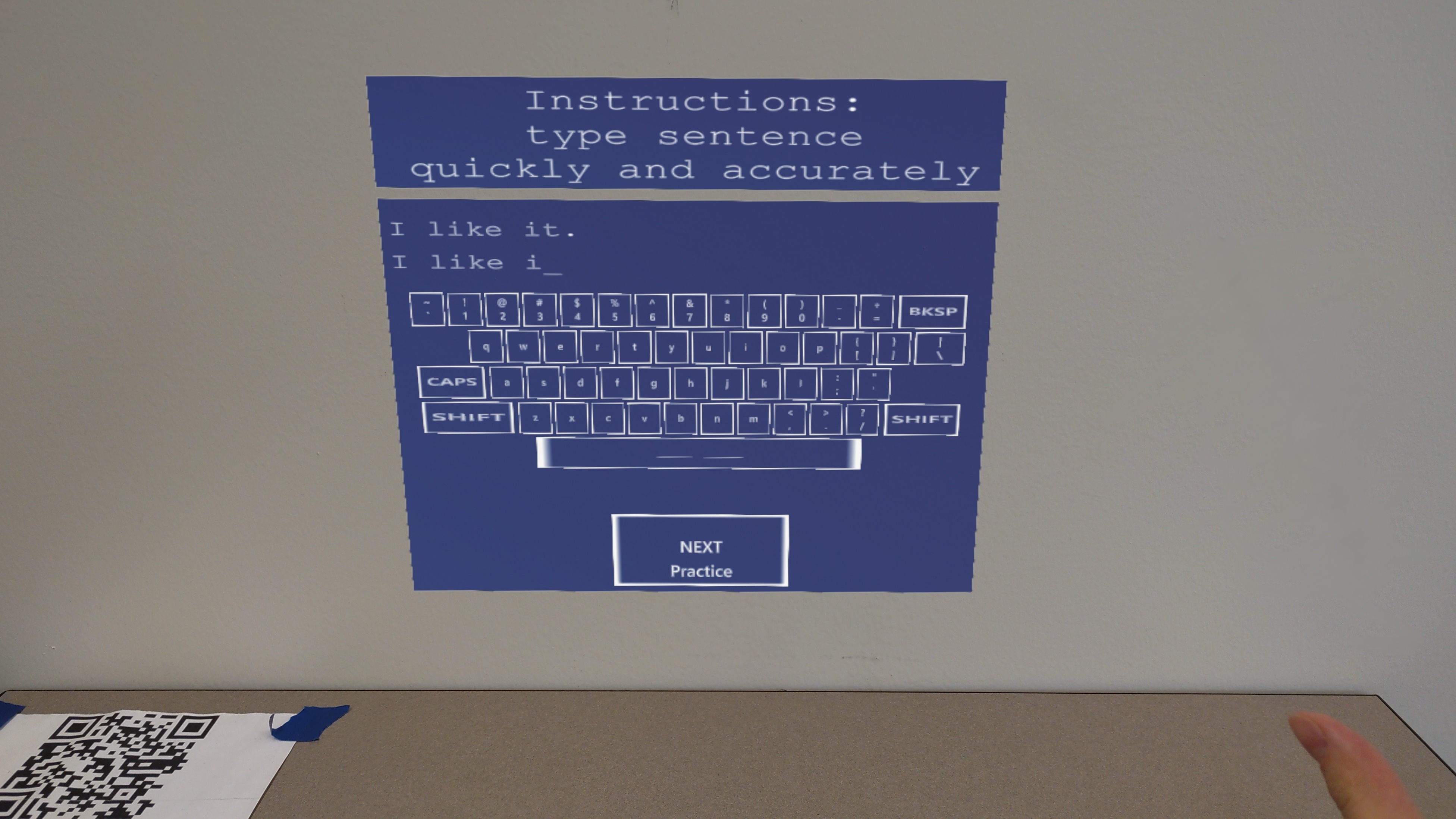}
\caption{The table the user was seated at for Experiment 1 and the HoloLens headset used (top left), the \textsc{Midair} condition (top right), the \textsc{Table} condition (bottom left), and the \textsc{Wall} condition (bottom right).}
\label{fig_exp1_screenshots}
\Description{A small table with an office chair, a keyboard floating midair in front of the user, a keyboard placed on the horizontal table surface, and a keyboard placed on the wall behind the table.}
\end{figure}

The experiment consisted of three conditions (depicted in Figure \ref{fig_exp1_screenshots}): 
\begin{itemize}
    \item \textsc{Midair --- } The keyboard was placed in midair in front of the participant. The keyboard was horizontally centered above the table with the bottom of the spacebar key being 19.0\,cm from the front of the table and 4.5\,cm above the table. This position made it possible for users to rest their forearms against the edge of the table while typing if they so desired. The keyboard was angled 15 degrees above horizontal. In pilot testing, we found this tilt made it more comfortable to type while seated compared to a vertical or completely horizontal keyboard. A similar tilt was used in a study where participants typed on or above a surface while seated \cite{dudley2019strategies} (c.f.~Figure 5). 
    \item \textsc{Table --- } The keyboard was placed just above a table. The keyboard was horizontally centered on the table with the bottom of the spacebar key being 28.5\,cm from the front of the table. This position made it possible for users to rest their arms on the table as they typed if they so desired.
    \item \textsc{Wall --- } The keyboard was placed slightly in front of a wall. The keyboard was horizontally centered on the table with the bottom of the spacebar being 16.0\,cm above the table. This position coupled with the small size of the table made it possible for users to rest their arms on the table as they typed if they so desired.
\end{itemize}

We hypothesized the keyboards positions on surfaces (i.e.~\textsc{Wall} and \textsc{Table}) would perform better than the baseline condition (i.e.~\textsc{Midair}). We also hypothesized \textsc{Table} would be preferred due to its similarity to a standard physical keyboard. We investigated both the \textsc{Wall} and \textsc{Table} locations to assess typing on common real-world flat surfaces. In addition, the keyboard position and orientation relative to the headset is different between the \textsc{Wall} and \textsc{Table} conditions. This allowed us to investigate whether this impacted hand-tracking accuracy.

In all conditions, participants typed with their index fingers on one or both hands. This choice prioritizes ecological validity by allowing participants to type as they would in the real world. Our keyboard was deterministic and did not perform auto-correction or predict words. We did this to understand the raw performance of typing as might be required for text not easily predicted by a language model.

All conditions had audio and visual feedback for key presses. A key was triggered when either index finger entered the key's bounding area, pushed the key's face a certain depth, and exited the same key's bounding area. On a key press, we played a  sound effect and displayed a white border around the key. We displayed participants' hands using the Unity MR toolkit's default hand skin, an overlay of white and grey triangles. We displayed blue spheres on index fingertips and grey cubes at joints. 

The keyboard size (including four character rows plus the spacebar) was 33.0\,cm $\times$ 11.0\,cm. Each letter key was 2.2\,cm $\times$ 2.2\,cm. The keyboard was similar in size to a physical desktop keyboard. It is also similar to the virtual key sizes reported in past work, e.g.~2.2\,cm \cite{dudley_augmented}, 2.5\,cm \cite{dudley2019strategies}, and 1.9\,cm \cite{richardson_stegotype}. The keyboard was rendered slightly transparent blue with white key labels. We used a full QWERTY keyboard including numbers, symbols, shift keys, caps lock, and backspace. Users could see their current text above the keyboard and could correct errors via a backspace key. 

\subsection{Procedure}

Participants used a HoloLens 2 headset while seated at a small table (Figure \ref{fig_exp1_screenshots} left). The table was 91\,cm wide and 51\,cm deep and we placed a QR code at its top left corner. At the start of each condition, participants had to look at the QR code. This provided a spatial anchor to allow consistent placement of the virtual keyboard in each condition. We implemented our spatial anchoring using a MR toolkit extension (see our source code for details). 

The study lasted approximately one hour. Participants completed an initial questionnaire with demographic questions, questions about their text entry experience, and questions about their experience with VR/AR. Between conditions, participants took a two-minute break. During the break, we instructed participants to lift the HoloLens visor to reduce eyestrain. Participants then completed a questionnaire about the previous condition. Participants ranked on a 7-point Likert scale (1=strongly disagree, 7=strongly agree) whether they felt the condition allowed them to enter text 1) quickly, 2) accurately, and 3) physically easily. We also solicited freeform feedback on what they liked and disliked about the previous condition. After all conditions, participants completed a final questionnaire in which they ranked conditions from best to worst and explained their ranking. We also asked them what they thought needed improving in AR in general. For our exact questionnaires, see our supplementary materials.\ifanon
\else
\footnote{\url{https://osf.io/q4n72}} 
\fi
~Participants received a \$15 USD gift card.

The order of conditions was completely counterbalanced between participants. Prior to each condition, participants typed a calibration sentence to set the keyboard's height above the surface. If the keyboard was too far in front of the surface, the finger could actuate a key without tactile feedback. If it was too close to the surface, the tracked finger would hit the real surface before it actuated a virtual key---making it impossible to type. Calibration computed the average depth of nine key presses when the participants typed ``calibrate''. The participants were instructed to keep pushing the key down until they felt the surface. While the calibration sentence only affected the \textsc{Wall} and \textsc{Table} conditions, we retained it in the \textsc{Midair} condition for consistency. During calibration, participants used only one hand to avoid potential depth measurement issues caused by sensing the other hand. 

Following calibration, participants typed two practice sentences and twelve evaluation sentences. We only analyzed the evaluation sentences. Sentences were pulled randomly without replacement from the ``mem1-5'' set of the mobile Enron dataset \cite{vertanen_mobilehci2011}. These sentences were in mixed case and included punctuation and sometimes numbers. During the practice and evaluation sentences, we reminded users they could type with both hands. 

\subsubsection{Independent variables and statistical tests.} Experiment 1 had a single independent variable, namely the location of the keyboard (midair, table, or wall). We tested for statistical differences using a one-way repeated measures ANOVA. In cases were Mauchly's test indicated a violation of sphericity, we report Greenhouse-Geisser corrected tests. For significant omnibus tests, we test for differences between conditions via Bonferroni corrected post-hoc tests. For Likert ratings, we used a non-parametric Friedman's ANOVA test. For significant omnibus tests, we test for differences between conditions via a Bonferroni corrected non-parametric Wilcoxon signed-rank test.

\subsubsection{Participants and demographics.} We recruited 18 participants aged 18--26. 15 participants identified as male, 2 as female, and 1 as non-binary. 16 participants strongly agreed with the statement ``I consider myself a fluent speaker of English'', one participant agreed, and one participant strongly disagreed. We believe the participant who strongly disagreed had made an error when marking the scale. Eight participants reported augmented reality (AR) experience, including one participant who reported owning a HoloLens. The remaining seven participants reported limited AR experience, primarily with phone apps (e.g.~Pokémon Go\footnote{Pokémon Go is an AR game released in 2016 in which players locate, capture, train, and battle virtual creatures that appear in real-world locations.}). 

\subsection{Results}

\subsubsection{Entry rate.} Entry rate was measured in words-per-minute (WPM). As is common in text entry research, we considered a word to be five characters (including space)~\cite{wobbrock_measures}. We subtracted one from the number of characters to compensate for the unmeasured time of finding and typing the first key \cite{wobbrock_measures}. We started timing from when the first key was pressed. We ended timing when the next button was pressed since this includes the perceptual, cognitive, and motor overheads associated with confirming the last character. 

Entry rates were 12.2, 9.1, and 7.8\,WPM in the \textsc{Midair}, \textsc{Table}, and \textsc{Wall} conditions respectively (Figure \ref{fig_exp1_wpm_cer} left). This difference was significant (Table \ref{table_exp1} left). Post-hoc tests found  \textsc{Midair} was significantly faster than both \textsc{Table} and \textsc{Wall}. 

\begin{table*}[tb]
\caption{Results for Experiment 1 (top) and statistical test details (bottom). Results are formatted as: mean $\pm$ sd [min, max]\label{table_exp1}.}
\begin{center}\small
\begin{tabular}{ l l l l }
\toprule
                & Entry rate (WPM) &  Error rate  (CER \%) & Backspaces per final character  \\
\midrule
\textsc{Midair} & 12.2 $\pm$ 2.1 [8.8, 17.6]   & 0.95 $\pm$ 1.9 [0.0, 7.8]   & 0.119 $\pm$ 0.092 [0.017, 0.330] \\
\textsc{Table}  & \hspace{1.3mm}9.1 $\pm$ 2.2 [5.8, 13.1]    & 0.98 $\pm$ 2.1 [0.0, 9.2]   & 0.160 $\pm$ 0.151 [0.015, 0.612] \\
\textsc{Wall}   & \hspace{1.3mm}7.8 $\pm$ 1.3 [5.9, 11.2]    & 0.65 $\pm$ 1.2 [0.0, 4.4]   & 0.132 $\pm$ 0.091 [0.040, 0.349] \\
\midrule
Omnibus test     & $F_{2,34}$\,=\,$31.3$, $\eta^2_G = .48$, $p < .001$  & $F_{2,34}=.89$, $\eta^2_G = .007$, $p=.42$  &   $F_{1.4,24.2}=1.73$, $\eta^2_G = .02$, $p=.20$  \vspace{1mm} \\
Pairwise         & \textsc{Midair}\,$>$\,\textsc{Table}, $p$\,<\,$.001$    &  n/a & n/a \\
post-hoc         & \textsc{Midair} $>$ \textsc{Wall}, $p$\,<\,$.001$       &  & \\
tests            & \textsc{Table}\,$\approx$\,\textsc{Wall}, $p$\,=\,$.12$ &  & \\
\bottomrule
\end{tabular}
\end{center}
\end{table*}

\subsubsection{Error rate.} 
\begin{figure}[tb]
\centering
\includegraphics[width=4cm]{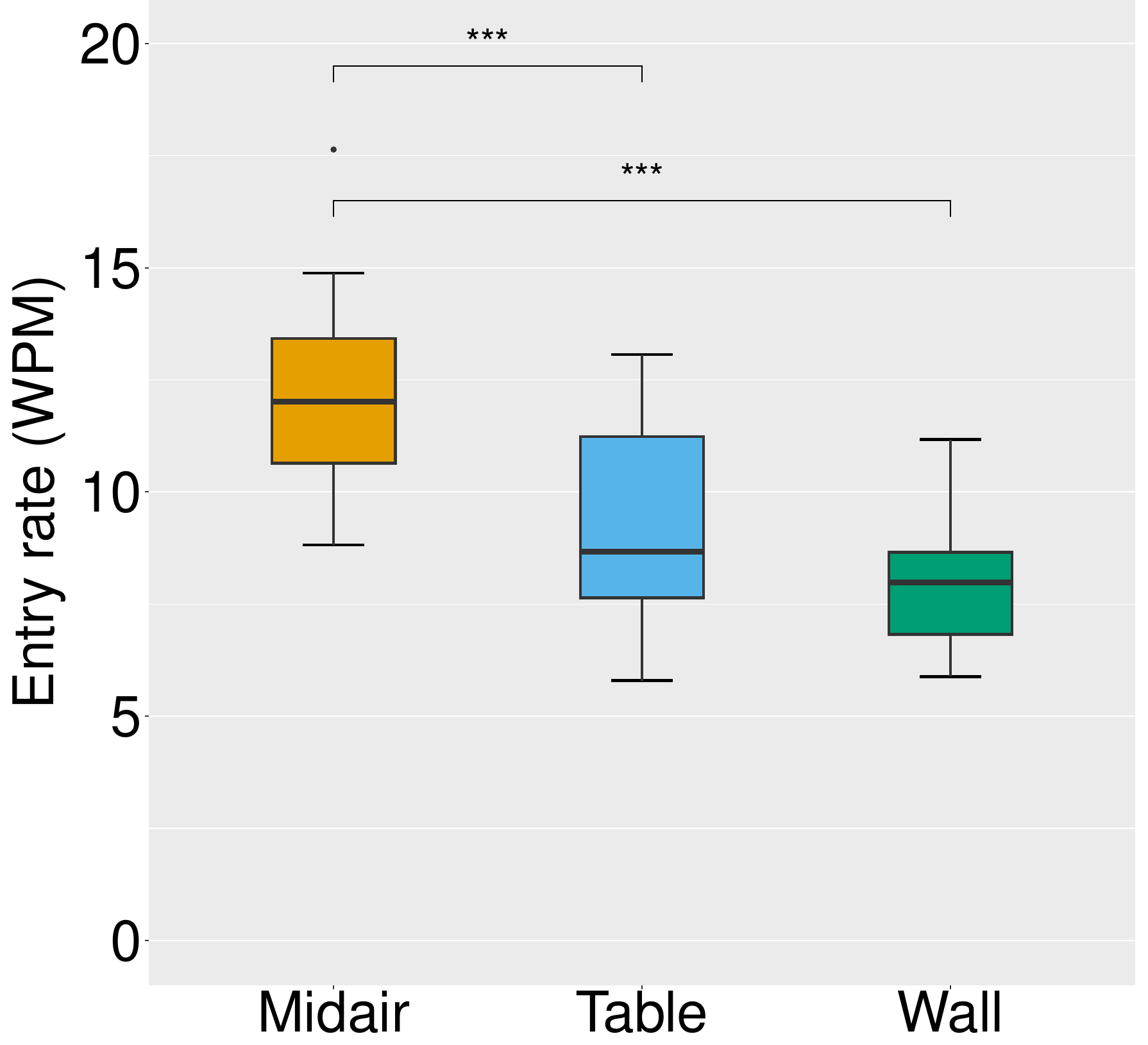} 
\hspace{5mm}
\includegraphics[width=4cm]{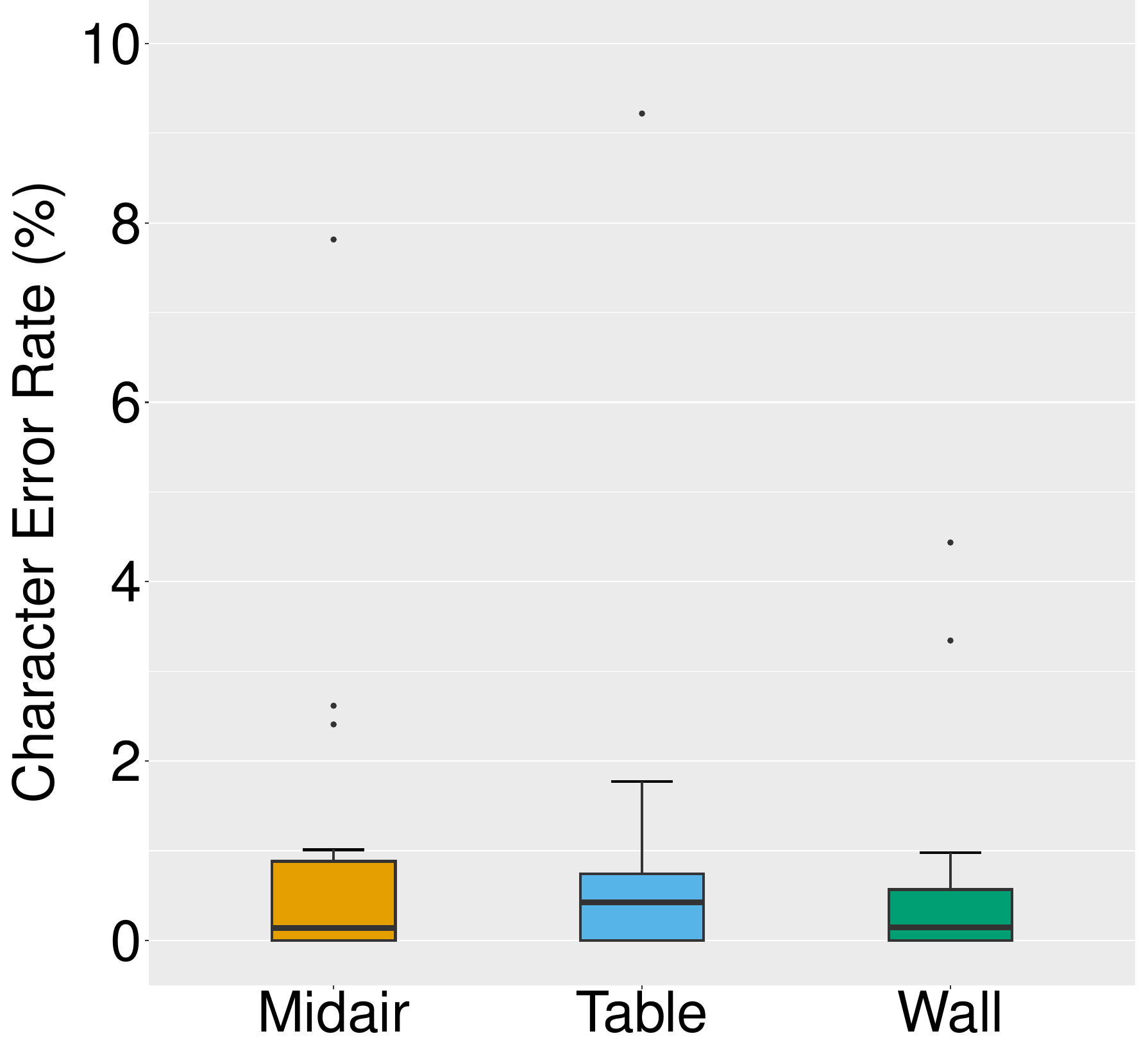}
\hspace{5mm}
\includegraphics[width=4cm]{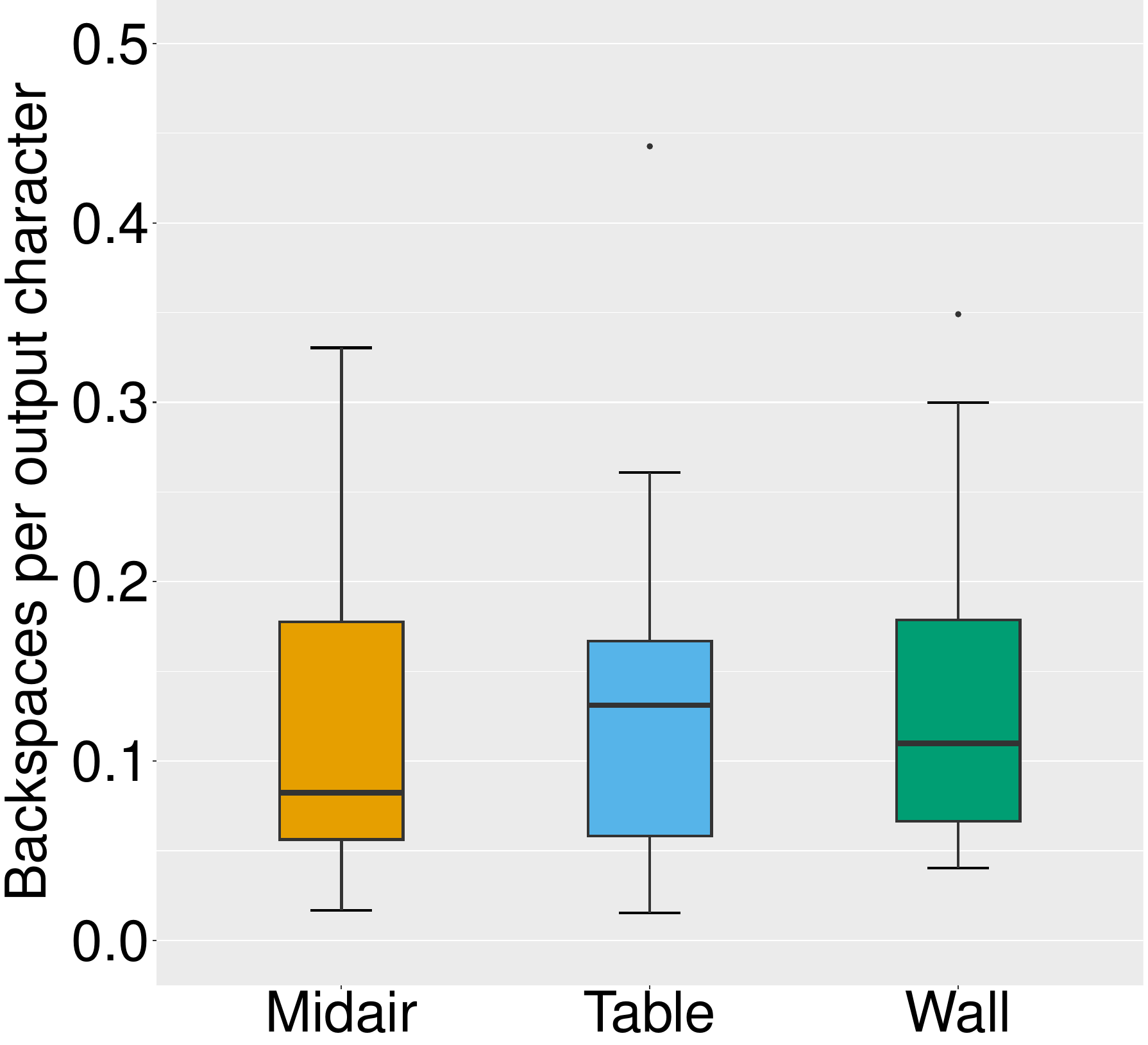}
\caption{Entry rate (left), error rate (center), and backspaces per final output character (right) in Experiment 1. Brackets with *** denote significant post-hoc pairwise differences at $p<0.001$.\label{fig_exp1_wpm_cer}}
\Description{Three box plots showing the entry rate, character error rate, and backspaces per character in Experiment 1. Each plot shows the results for each of the three conditions.}
\end{figure}

We measured typing errors using character error rate (CER). CER refers to the number of character insertions, substitutions, and deletions required to change the final entered text to the reference text, relative to the characters in the reference and multiplied by 100. Error rates were low in all conditions with a CER of 0.95\%, 0.98\%, and 0.65\% in \textsc{Midair}, \textsc{Table}, and \textsc{Wall} respectively (Figure \ref{fig_exp1_wpm_cer} center). This difference was not significant (Table \ref{table_exp1} center).

\begin{figure}[tb]
\centering
\includegraphics[height=4.5cm]{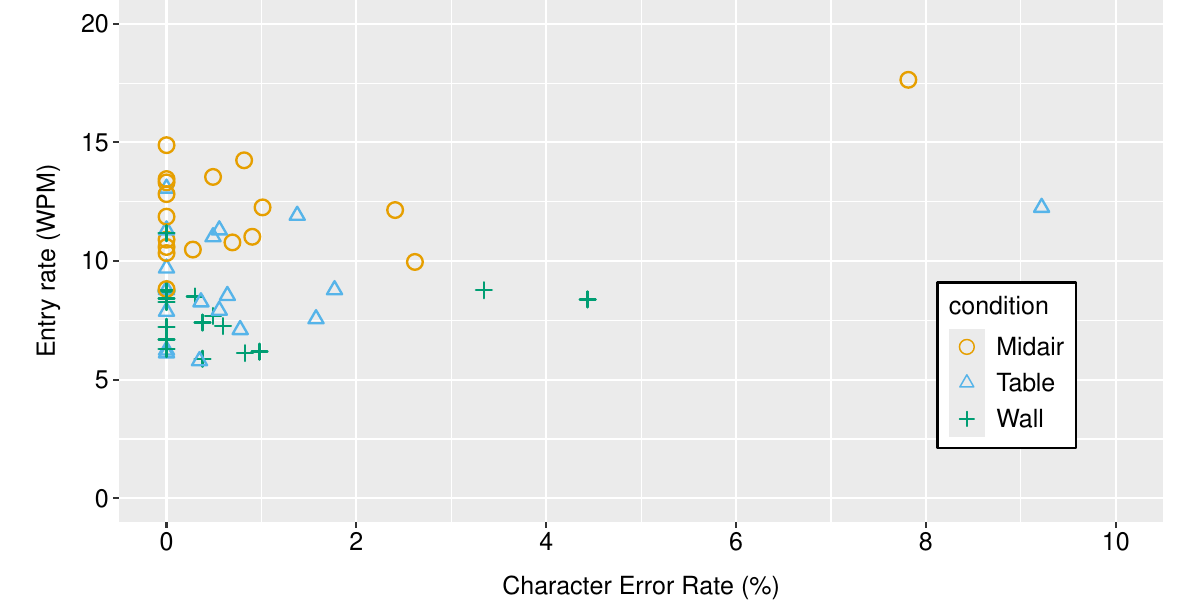} 
\caption{Entry and error rate of each participant in each condition in Experiment 1.\label{fig_exp1_scatter}}
\Description{Scatter plot showing how each participant did in each condition of Experiment 1 with respect to the character error rate (x-axis) and entry rate (y-axis).}
\end{figure}

We were curious how performance differed among participants. Figure \ref{fig_exp1_scatter} shows each participant's performance in each condition. In \textsc{Midair}, except for two participants, everyone wrote at 10\,WPM or faster with a CER of less than 3\%. In all conditions, with the exception of two participants, everyone obtained a final error rate of less than 5\%. In all conditions, many participants corrected all errors to achieve exactly the requested text (including capitalization and punctuation). This demonstrates typing on a virtual keyboard is a viable, albeit somewhat slow, mixed reality text entry method. This includes for text that cannot be auto-corrected (e.g.~passwords, account numbers, and URLs).

\subsubsection{Correction rate.} As a measure of how much effort participants were expending correcting errors, we calculated the number of backspaces per final output character for each phrase. While the final character error rates in the conditions were relatively low, we observed frequent use of backspace. The number of backspaces pressed per output character was 0.12 for \textsc{Midair}, 0.16 for \textsc{Table}, and 0.13 for \textsc{Wall}. While there did seem to be more backspacing in the two surface keyboards (in particular for \textsc{Table}), this difference was not significant (Table \ref{table_exp1} right).
% ($F_{2,34}=1.73, p=0.19)$.

\subsubsection{Hand-tracking success.} We recorded participants' left and right index finger positions 30 times per second, including noting whether tracking was lost for either finger. We computed the percentage of samples where at least one finger was tracked. Participants' tracking success percentages during the evaluation phrase were 99.9\%, 99.1\%, and 98.6\% in \textsc{Midair}, \textsc{Table}, and \textsc{Wall} respectively. While this difference was significant ($F_{2,34}$\,=\,$4.23$, $\eta^2_G = 0.10$, $p < 0.05$), post-hoc pairwise comparisons were not significant: 
\textsc{Midair}\,$\approx$\,\textsc{Table} ($p$\,=\,$0.26$), 
\textsc{Midair}\,$\approx$\,\textsc{Wall} ($p$\,=\,$0.07$), and  
\textsc{Table}\,$\approx$\,\textsc{Wall} ($p$\,=\,$0.69$).
We note that just because we tracked both fingers, this does not mean the returned location was necessarily accurate. As we will discuss, subjective feedback from many participants indicated hand-tracking was less accurate in \textsc{Midair} and \textsc{Table}. We conjecture this could have resulted from a number of factors such as the proximity of the hands to the surface, occlusion of the fingers or hands between themselves, or changes due to the relative location and angle between the hands and the headset.

\begin{figure}[tb]
\subfloat[\textsc{Midair}\label{fig_keys_midair}]{%
  \includegraphics[width=7.3cm]{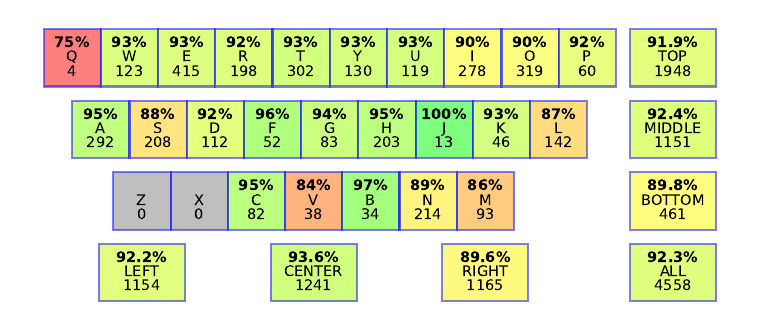}
}
\subfloat[\textsc{Table}\label{fig_keys_table}]{%
  \includegraphics[width=7.3cm]{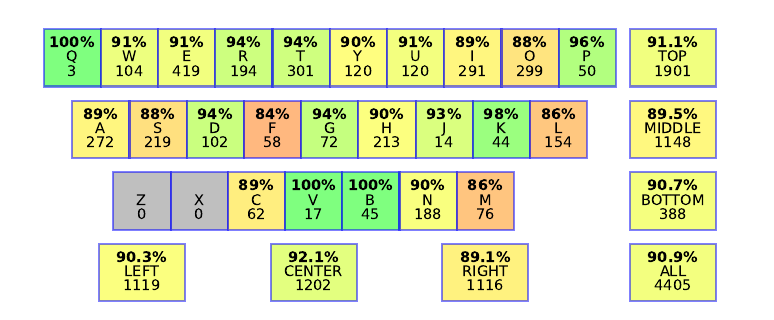}
}\vspace{-2mm} \\
\subfloat[\textsc{Wall}\label{fig_keys_wall}]{%
  \includegraphics[width=7.3cm]{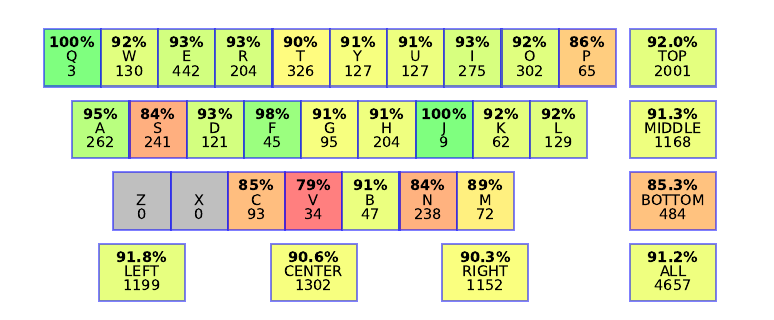}
}
\caption{Accuracy of key press attempts for each letter key in each condition. The integer at the bottom of each key is how many events the percentage was calculated over. We also grouped attempts by vertical keyboard row (i.e.~top, middle, bottom) and by horizontal area (i.e.~left, center, right). For the top and middle rows, we considered the three leftmost and rightmost keys as the left and right areas respectively. For the bottom row, we considered the two leftmost and rightmost keys as the left and right areas respectively.}
\label{fig_keys}
\Description{Three figures of a QWERTY keyboard. Each letter is labeled with the percentage of attempts to type that letter that were successful.}
\end{figure}

\subsubsection{Accuracy over keyboard.} To investigate if accuracy was impacted by the spatial location of a key, we calculated the percentage of time participants successfully obtained the next letter in the target phrase. We calculated this only on key presses in which, at the time of the key press, a participant's current typing result was a strict prefix of the target. For example, if the target phrase was ``Are you there?'' and the participant had typed ``Are you t'' we scored the next key press based on a target of ``h''. However if they had typed ``Are you h'' we would drop the next key press as we are unsure if they missed the ``t'' key or if they are writing an incorrect word like ``here''. This filtering allowed us to be reasonably confident about a participant's next intended key. As shown in Figure \ref{fig_keys}, there did not seem to be a strong difference in accuracy based on spatial location in \textsc{Midair} or \textsc{Table}. However in the \textsc{Wall} condition, the bottom keyboard row was noticeably less accurate at 85\% versus 91\% for the middle row and 92\% for the top row.

\subsection{Subjective feedback}
\label{exp1-subjective-feedback}

\begin{figure}[tb]
\includegraphics[width=15cm]{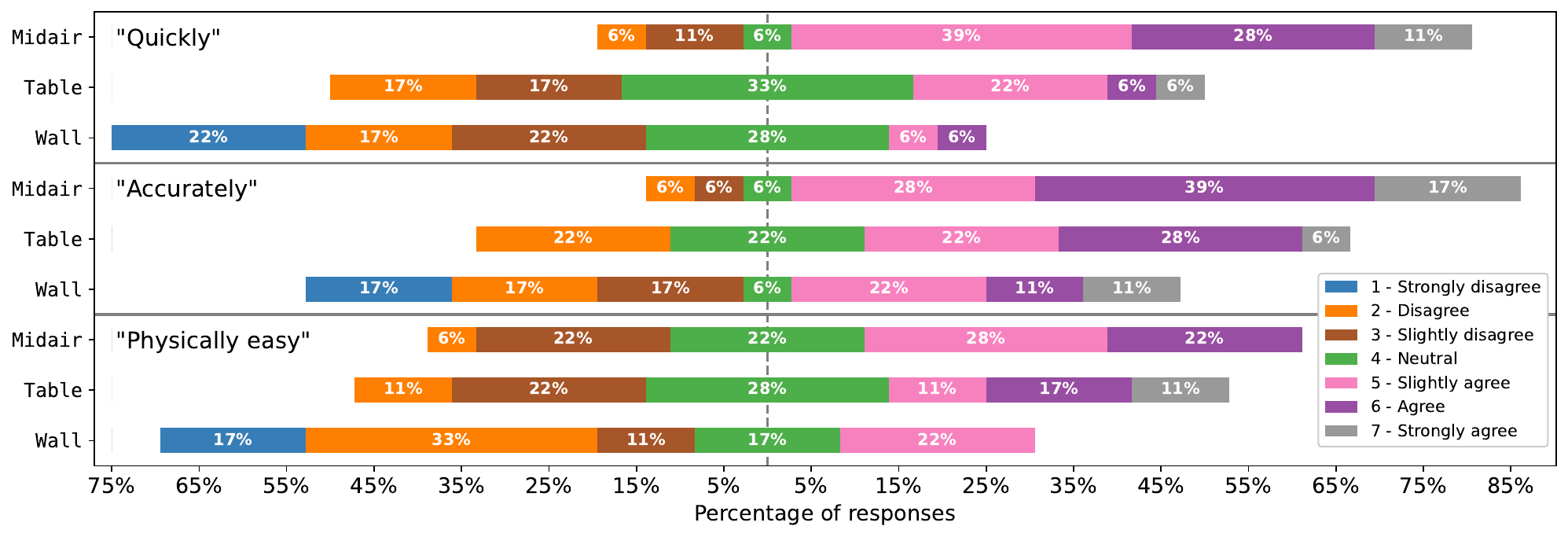}
\caption{Participants' Likert responses in Experiment 1 to the statements ``I entered text quickly'' (top), ``I entered text accurately'' (middle), and ``Entering text was physically easy and not tiring'' (bottom).}
\label{fig_exp1_likert}
\Description{Image showing the Likert ratings for three statements in Experiment 1.}
\end{figure}

\subsubsection{Likert statements.} After each condition, participants rated three statements on a 7-point Likert scale (Figure \ref{fig_exp1_likert}).
Participants' average ratings for ``I entered text quickly'' were: \textsc{Table} 4.0, \textsc{Wall} 2.9, and \textsc{Midair} 5.1. The difference was significant ($\chi^2(2)=25.9, p<0.001$). The only significant post-hoc test was that \textsc{Midair} was quicker than \textsc{Wall} ($p<0.001$). 
Participants' average ratings for ``I entered text accurately'' were:  \textsc{Table} 4.5, \textsc{Wall} 3.8, and \textsc{Midair} 5.4. The difference between ratings was significant ($\chi^2(2)=9.2, p<0.01$). The only significant post-hoc test was that \textsc{Midair} was more accurate than \textsc{Wall} ($p<0.05$). 
Participants' average ratings for ``Entering text was physically easy and not tiring'' were: \textsc{Table} 4.3 for, \textsc{Wall} 2.9, and \textsc{Midair} 4.4. The difference between ratings was significant ($\chi^2(2)=10.9, p<0.01$). The only significant post-hoc test was that \textsc{Midair} was physically easier than \textsc{Wall} ($p<0.05$). 

\subsubsection{Condition ranking.} We asked participants to rank the conditions and to explain their reasoning. Of the 18 participants, 13 preferred \textsc{Midair}, 5 preferred \textsc{Table}, and none preferred \textsc{Wall}. We asked participants to explain their ranking of the conditions. Of those who ranked \textsc{Midair} the best, some representative comments were: ``midair was the fastest and tracked most accurately'', and ``hovering was most comfortable and easy to use''. Of those ranking \textsc{Table} as the best, participants commented: ``horizontal was natural feeling and responsive'', and ``horizontal provided the most comfort and accuracy for me''. Most participants ranked \textsc{Wall} as the worst and participants often mentioned physical strain: ``vertical was the slowest and most physically taxing'', and ``wall was neither comfortable nor accurate''.

\subsubsection{Likes and dislikes.} We asked participants what they most liked and most disliked about the system after each condition. One author performed qualitative coding on these responses and another peer reviewed the codings and made minor improvements. For all participant comments, see our supplementary materials.\ifanon
~
\else
\footnote{\url{https://osf.io/q4n72}} 
\fi

Fatigue was mentioned by 8, 6, and 2 participants in the \textsc{Wall}, \textsc{Midair}, and \textsc{Table} conditions respectively. There was mixed feedback about the tactile feedback. 8 participants said they liked the tactile feedback in the \textsc{Wall} condition and 4 liked it in the \textsc{Table} condition. For example, one participant liked that they could ``touch the wall instead of nothing'' and another liked the ``physical sensation of touching the table.'' None complained about the the presence of haptic feedback in those conditions. However, in the \textsc{Midair} condition, three people reported liking no tactile feedback and one disliked not having tactile feedback. 

We counted people who mentioned poor tracking in some form (unexpected key presses, failing to detect key presses, or poor tracking in general). 9 people reported poor tracking in \textsc{Wall}, 5 reported it in \textsc{Midair}, and 13 people reported it in \textsc{Table}. Comments disliking tracking in the \textsc{Table} condition included ``it seemed I lost tracking a lot in this part'' and ``sometimes I would hit one key to the left or right of the intended key.''

\subsubsection{AR improvements.} Finally, we asked participants ``what needs to be improved or changed about the AR headset's hardware or software before you would be willing to use it in place of your mobile phone?'' Requested changes were: 
\begin{itemize}
    \item Improved hand-tracking accuracy and responsiveness (10 participants).
    \item Interaction with multiple fingers (10 participants).
    \item Improvements related to the visual capabilities of the headset (6 participants).
    \item Auto-correct or word predictions (3 participants).
    \item Ability to locate the keyboard wherever you wanted it (1 participant).
    \item Swipe (word-gesture keyboard) typing (1 participant).
    \item Smaller form factor headset (1 participant).
\end{itemize}

\subsection{Discussion}

Our goal in Experiment 1 was to assess whether a keyboard positioned on a surface outperformed or was preferred compared to typing in midair. We conjectured that the keyboards located on surfaces, especially the \textsc{Table} condition which is similar to a standard physical keyboard, would perform better and be more preferred. However, our data showed the \textsc{Midair} condition had a significantly higher entry rate than both \textsc{Table} and \textsc{Wall}. All conditions had low final error rates, though moderate use of backspace was observed. Furthermore, 13 of the 18 participants preferred the \textsc{Midair} condition, while the remaining 5 preferred the \textsc{Table} condition. Although \textsc{Wall} was ranked last, it was also the condition with the most people mentioning how they liked the tactile feedback. We observed throughout the study that the \textsc{Table} and \textsc{Wall} conditions reduced the reliability of key actuation and this was corroborated by participant feedback. This may explain the increased performance and preference for the \textsc{Midair} condition.

Poor performance in the \textsc{Table} condition may be due to our calibration process or the hand-tracking. During calibration, the participant typed the word ``calibrate''. We instructed participants to continue pushing the key until their finger touched the surface. This form of calibration was dependent on how precise the participant was. We found that with quick hand movements, the HoloLens' hand-tracking lagged and when the hand abruptly stopped, the resulting depth measurement could be deeper than the participant's finger had actually gone. The latency of the hand-tracking may also have caused inaccuracies, \citet{kim_star} report a latency of 90\,ms and tracking jitter of 1\,mm using a HoloLens 2 headset. It might be possible to improve the keyboard's placement above the surface with a modified or extended calibration procedure. It might also be advisable to allow users to interactively adjust the keyboard height during typing. A different tracking system may have produced different results. However, this work provides valuable information about how well position a keyboard on a surface works given the tracking of a commodity MR headset.

We also speculate the unreliable key triggering were related to limitations of the HoloLens' egocentric cameras or its hand-tracking models. During the experiment, we saw participants try to compensate for hand-tracking issues in a number of ways, such as moving closer to the keyboard or adjusting the viewing angle between the headset and the keyboard. However, in some cases, such adjustments only made tracking less reliable. Ideally, the HoloLens interface would provide some sort of informative feedback that could guide users to adjust their position or environment to improve tracking. Presently, the only feedback is the unstable rendering of a user's hand skeleton.

Midair interaction can be tiring since users have to hold their hands up with little support \cite{boring2009scroll,hincapie2014consumed}. While we positioned our midair keyboard such that participants could brace their forearms against the table edge, the like/dislike qualitative feedback suggested that the \textsc{Midair} condition was nearly as fatiguing as the \textsc{Wall} condition. However, we were surprised to find participants rated \textsc{Midair} the least tiring in the Likert responses (Figure \ref{fig_exp1_likert} bottom). The Likert and like/dislike data both show that \textsc{Table} was the most comfortable, likely because people could rest their arms on the table similar to traditional keyboard typing. Both sets of data also show that the \textsc{Wall} condition was the most physically demanding. Although participants could rest their elbows on the table in this condition, it might have worked better for some participants than others. This condition is also the most different from traditional typing and placed the keyboard the furthest away from the user. For consistency with the other conditions, participants stayed seated at the table while typing on the wall. Wall typing might be more ergonomic while standing. 

\section{Experiment 2}
Experiment 1 showed a significantly higher entry rate for the midair keyboard and users strongly preferred that keyboard. The majority of participants also requested the ability to type with multiple fingers. In Experiment 2, we focus exclusively on a midair typing. Our goals were to 1) explore if ten-finger typing is feasible on a midair keyboard, and 2) determine if eye-tracking could aid in eliminating accidental key presses.

\subsection{Design}

In pilot testing of ten-finger typing, we found users had difficulty maintaining a consistent hand position in midair. Monitoring the positions of all ten fingers was difficult and this resulted in a high number of accidental key presses. The lack of tactile feedback also meant users often did not realize they had unintentionally pressed a key. Because of this, we worried that ten-finger typing, on its own, would not be competitive with index-finger typing.

Increasingly, VR and AR headsets (including the HoloLens 2) have built-in eye-tracking. We hypothesized users would likely be looking at or near the key that they are trying to press. By tracking users' eye gaze locations, we could filter out key presses that were sufficiently far from where they are looking. We hoped that this would improve efficiency by reducing the time spent carefully monitoring hand positions and correcting accidental key presses.

\begin{figure}[tb]
    \centering
    \includegraphics[width=8cm]{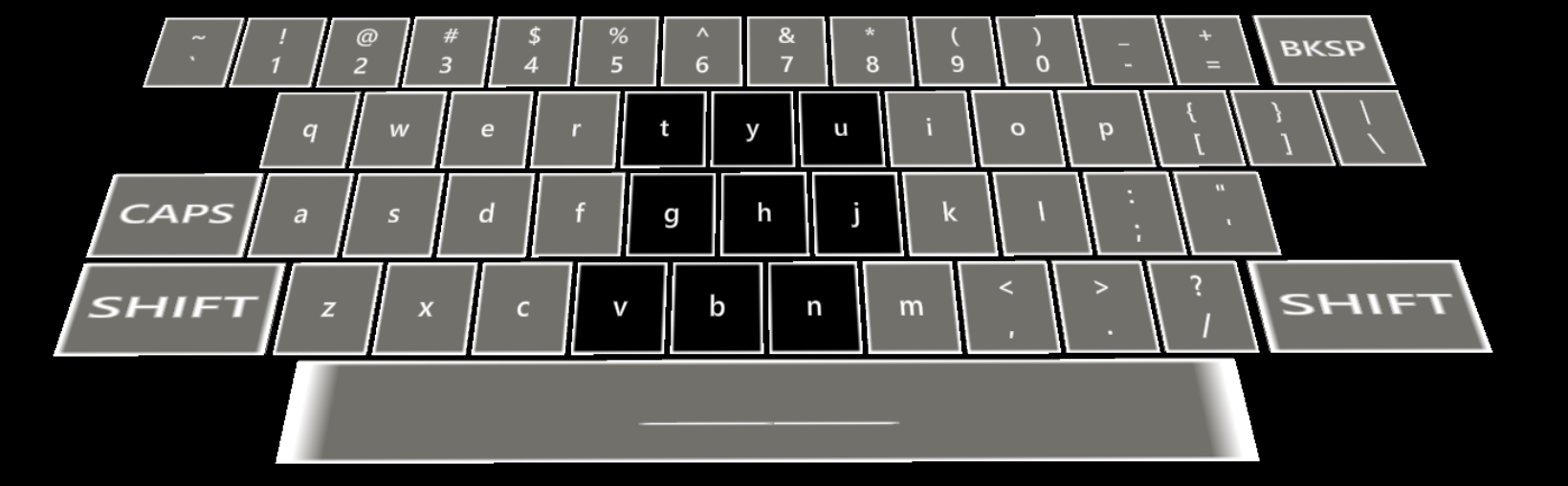}
    \caption{During the \textsc{IndexEye} and \textsc{TenEye} conditions in Experiment 2, disabled keys were grayed out and only the keys near a user's gaze position appeared with a transparent background. In this example, the user is looking at the ``h'' key. \label{fig_gaze_highlight}}
    \Description{Image of a full QWERTY keyboard. All the keys have a grey background except for the H key and the keys adjacent to it which all have a transparent background.}
\end{figure}

Like Experiment 1, the keyboard was deterministic without auto-correct or word predictions. In the conditions using the eye-tracking filtering, we enabled all keys within a 2.6\,cm radius of the user's gaze location. This effectively enabled the key the user was looking at plus all adjacent keys (Figure \ref{fig_gaze_highlight}). Keys outside this radius were disabled. Disabled keys were gray, while enabled keys were transparent. When pressed, a key flashed a slightly transparent blue and a sound was played. All keyboard text was white.

Experiment 2 consisted of four conditions:
\begin{itemize}
    \item \textsc{Index --- } We asked participants to type only with their index fingers. Key presses from other fingers were ignored.
    \item \textsc{IndexEye --- }  We asked participants to type only with their index fingers and to look at the keys as they typed. Key presses from other fingers were ignored as were keys outside of the radius around their gaze location.
    \item \textsc{Ten --- }   We asked participants to type with all ten fingers. Key presses were triggered by all fingers.
    \item \textsc{TenEye --- } We asked participants to type with all ten fingers and to look at the keys as they typed. Key presses were registered for all fingers. Keys outside of the radius around their gaze location were ignored.
\end{itemize}

\begin{figure}[tb]
    \centering
    \includegraphics[height=3.6cm]{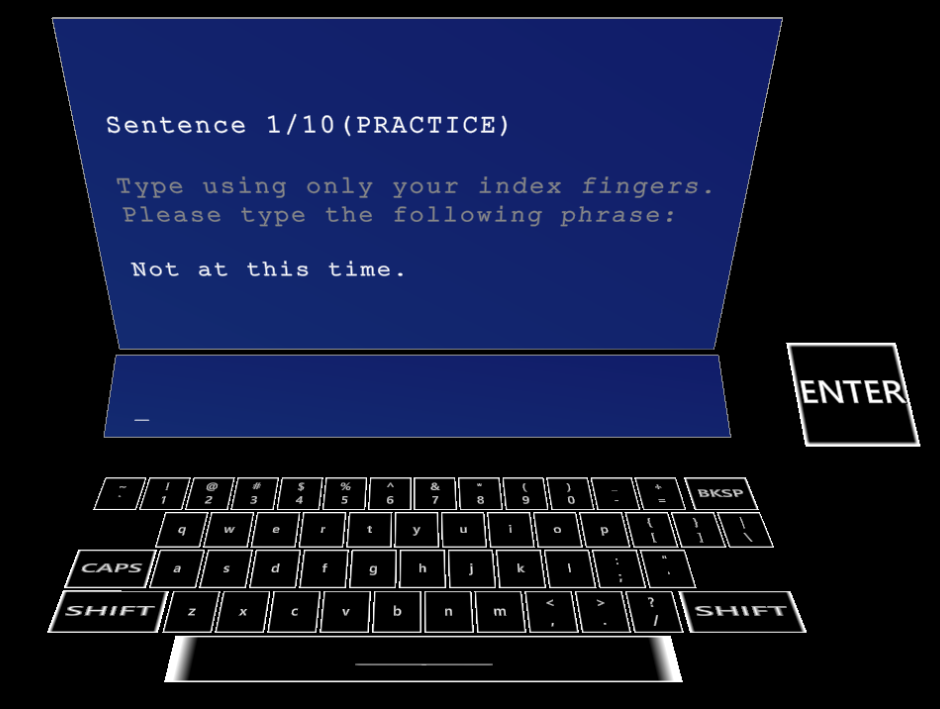} \hspace{1mm}
    \includegraphics[height=3.6cm]{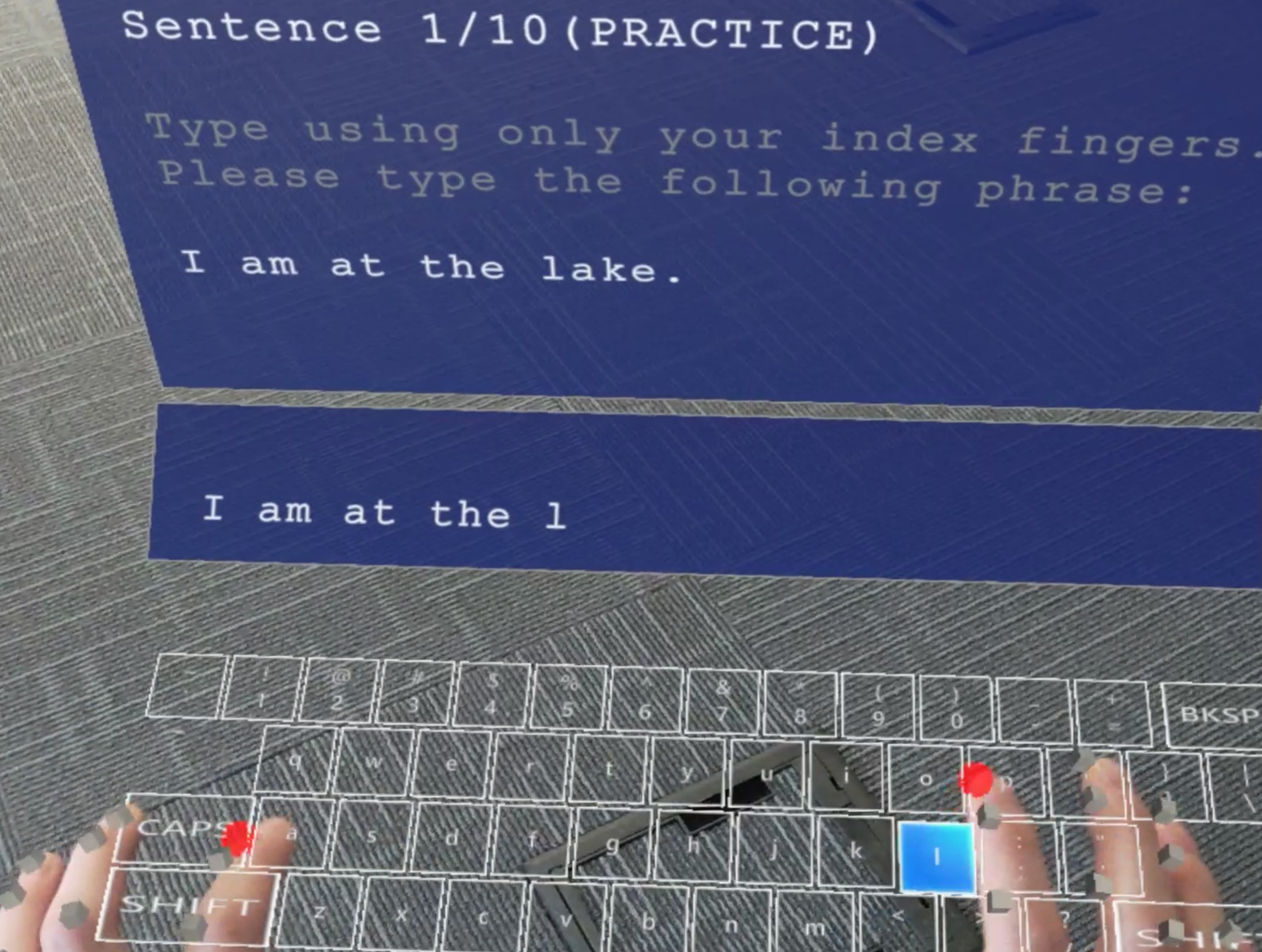} \hspace{1mm}
    \includegraphics[height=3.6cm]{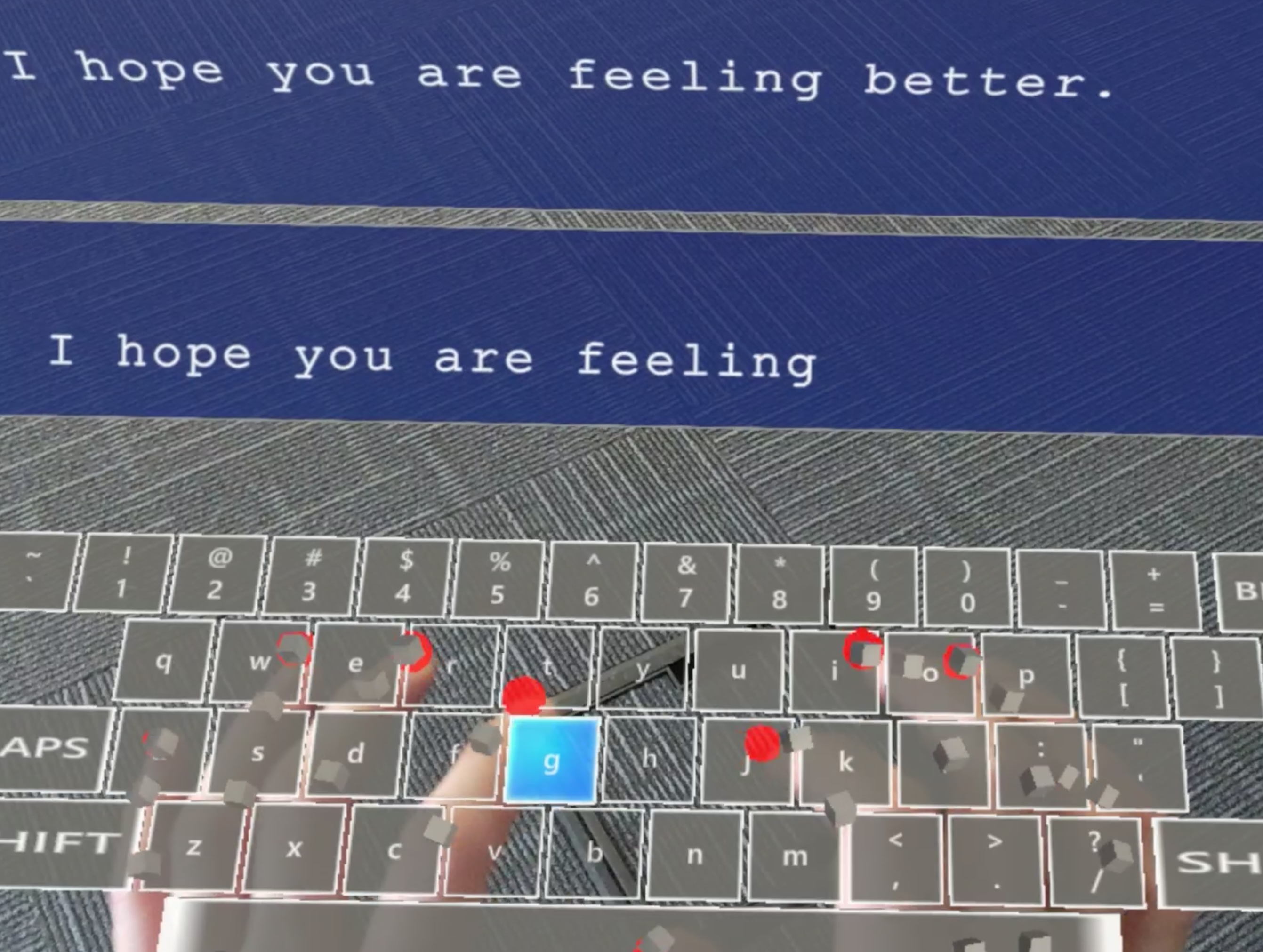}
    \caption{Overview of the keyboard and prompt (left), the \textsc{Index} condition (center), and the \textsc{TenEye} condition (right) in Experiment 2.}
    \label{fig_keyboard}
    \Description{The first image shows the interface used in Experiment 2. It consists of a keyboard, a middle area where the user's typed text appears, and an upper area that displays the phrase the user is copying. The second image shows a user typing with the index fingers of both hands. Each index fingertip has a red sphere on it. The third image shows a user typing with all ten fingers. Each fingertip has a red sphere on it.}
\end{figure}

A key was triggered when a participant's finger pushed a key downward past a threshold of approximately one-third of a centimeter. In the event that multiple keys were pressed at the same time, only the furthest pressed key was registered as pressed. Following a key press, there was a half-second cooldown where the keyboard would not register any key presses to help prevent multiple keys from being triggered from a single tap gesture. 

We used a QWERTY keyboard including numbers, symbols, shift keys, caps lock, and backspace. The keyboard size (including four rows of characters plus spacebar) was 35\,cm\,$\times$\,12\,cm. Each letter key was 2.2\,cm\,$\times$\,2.2\,cm. Red spherical colliders were attached to a user's index fingers in \textsc{Index} and \textsc{IndexEye}, and all ten fingers in \textsc{Ten} and \textsc{TenEye}.

We opted for a visual design mimicking a laptop (Figure \ref{fig_keyboard}) since there was no need for the interface to appear in a single plane (as was the case to support the \textsc{Wall} condition in Experiment 1). The top virtual screen shows the phrase the participant is being asked to type. The center text area shows the current text typed thus far. Typing occurs on the lower keyboard area, with the end of a sentence signaled via the large enter key on the right. A backspace key allows deletion of previously entered characters. As in Experiment 1, the keyboard was tilted 15 degrees above horizontal.

\subsection{Procedure}
The study took approximately one hour. Participants first completed an initial questionnaire. They then donned the HoloLens 2 headset and calibrated the eye tracker before proceeding to their first condition. The keyboard was placed midair in front of the participant with the bottom of the spacebar located 40\,cm below eye level. In pilot testing, we determined this to be a comfortable seated typing location for most people. Participants were instructed to adjust the chair height and location to fine-tune their position with respect to the keyboard. Since the keyboard was intended to be used in midair only, no table was used. Participants were instructed to type using both hands. 

Between conditions, participants took a two-minute break and completed a questionnaire about the condition just finished. During this time, we instructed participants to remove the headset and walk around the room to reduce eyestrain. After all conditions, participants completed a final questionnaire. The questionnaires were identical to Experiment 1 and are included in our supplementary materials.\ifanon
\else
\footnote{\url{https://osf.io/q4n72}}
\fi
~Participants received a \$15 USD gift card. 

The order of conditions was completely counterbalanced between participants. Each condition consisted of two practice sentences, followed by eight evaluation sentences. We only analyzed the evaluation sentences. For each condition, sentences were pulled randomly without replacement from the ``mem1-5'' set from the mobile Enron dataset~\cite{vertanen_mobilehci2011}. These sentences were in mixed case and included punctuation and sometimes numbers. 

\subsubsection{Independent variables and statistical tests.} We tested for statistical differences using a two-way repeated measures ANOVA. Our two independent variables were: 1) whether eye-tracking filtering was enabled (denoted \textit{Eye}), and 2) whether participants typed with index fingers or all ten fingers (denoted \textit{Fingers}). In the case of significant interaction effects, we tested for differences between conditions via Bonferroni corrected post-hoc tests. For Likert ratings, we used a non-parametric aligned rank transform \cite{wobbrock_aligned}.

\subsubsection{Participants and demographics.} We recruited 24 participants aged 18--22. Of these, 17 were male, 5 were female, and 2 preferred not to say. 23 participants strongly agreed with ``I consider myself a fluent speaker of English'', and the remaining participant agreed with the statement. A total of seven participants reported AR experience, with four participants having used a HoloLens before, and three participants reporting use of phone AR apps (e.g.~Pokémon Go).

\subsection{Results}
\subsubsection{Entry rate.} 
\begin{figure}[tb]
\centering
\includegraphics[width=4cm]{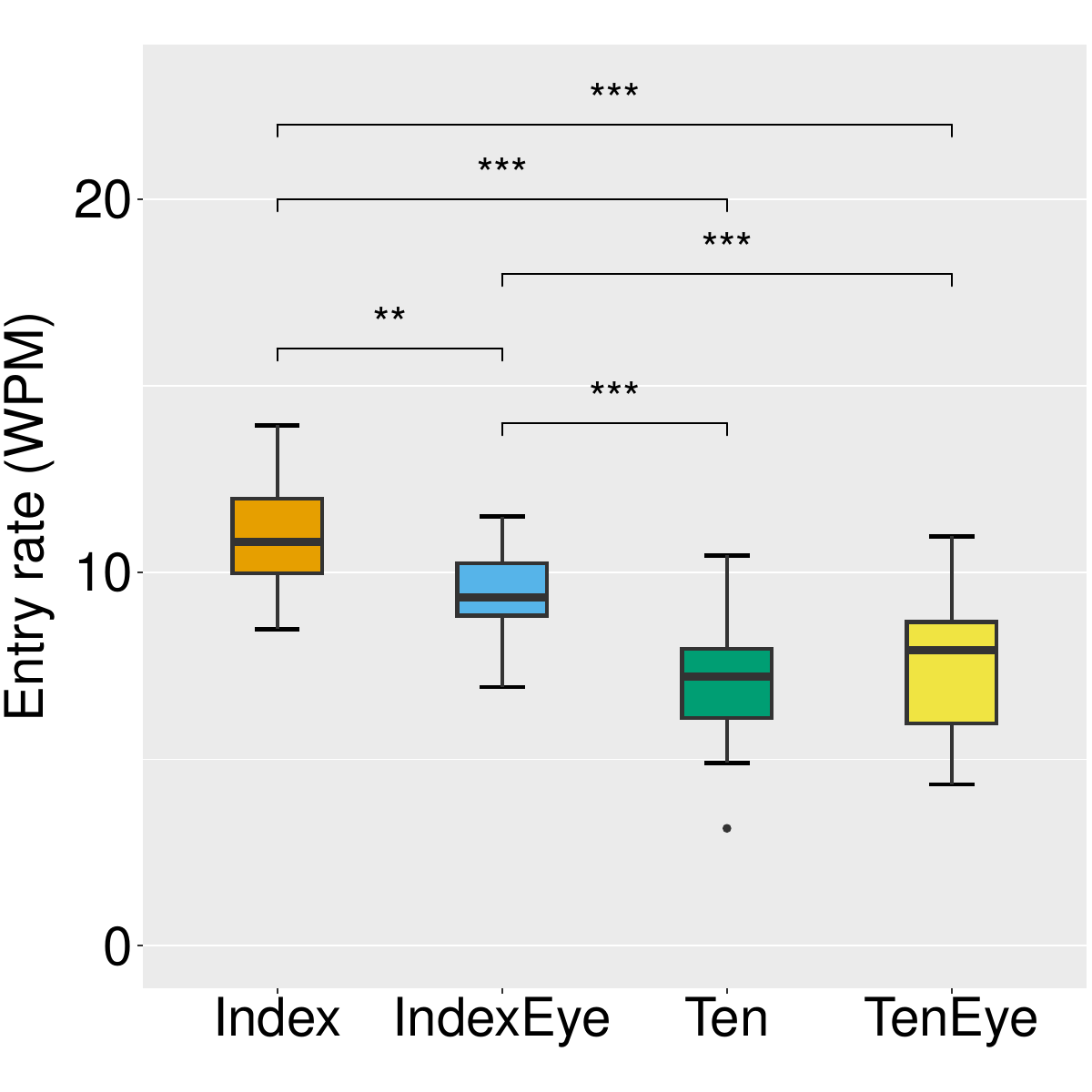} 
\hspace{5mm}
\includegraphics[width=4cm]{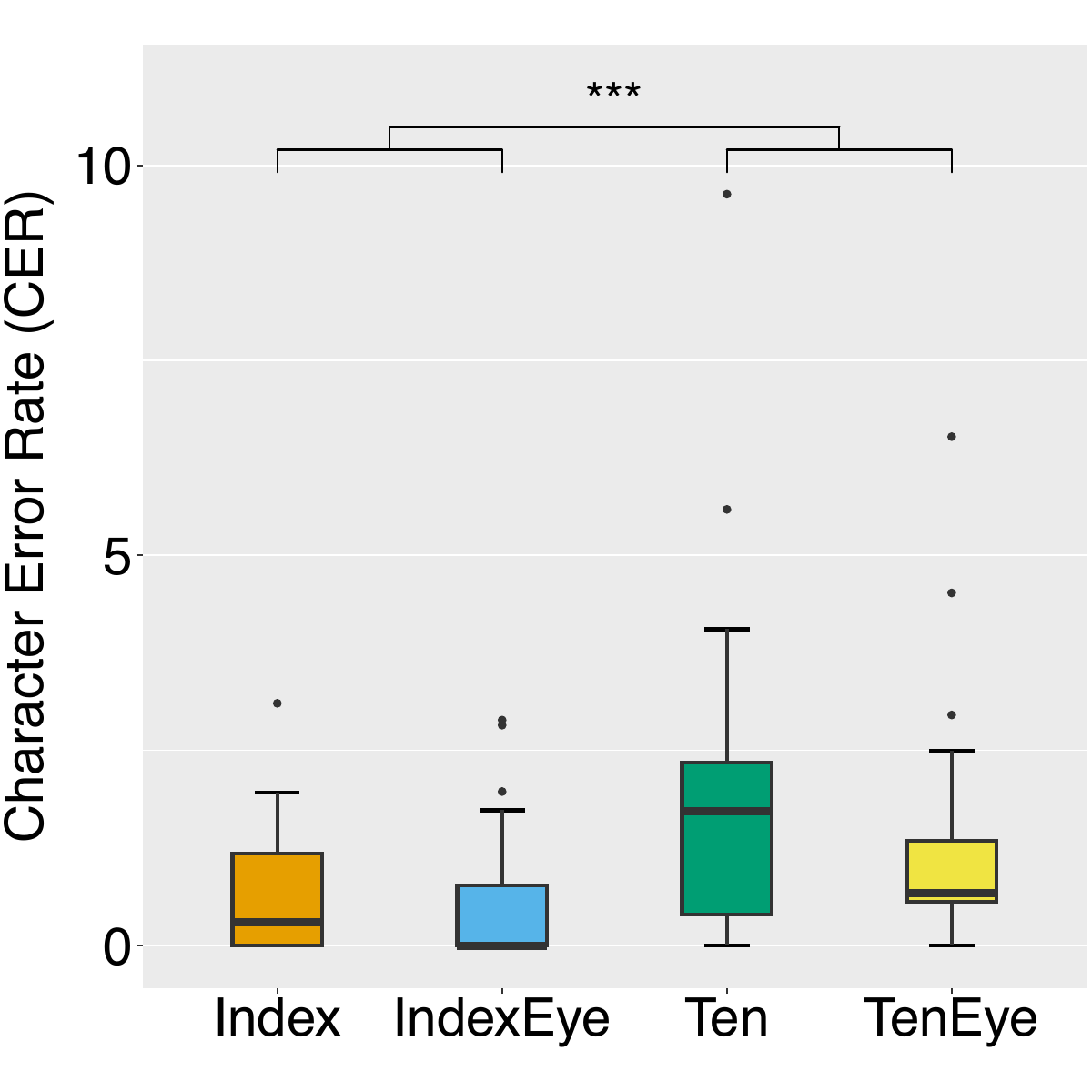}
\hspace{5mm}
\includegraphics[width=4cm]{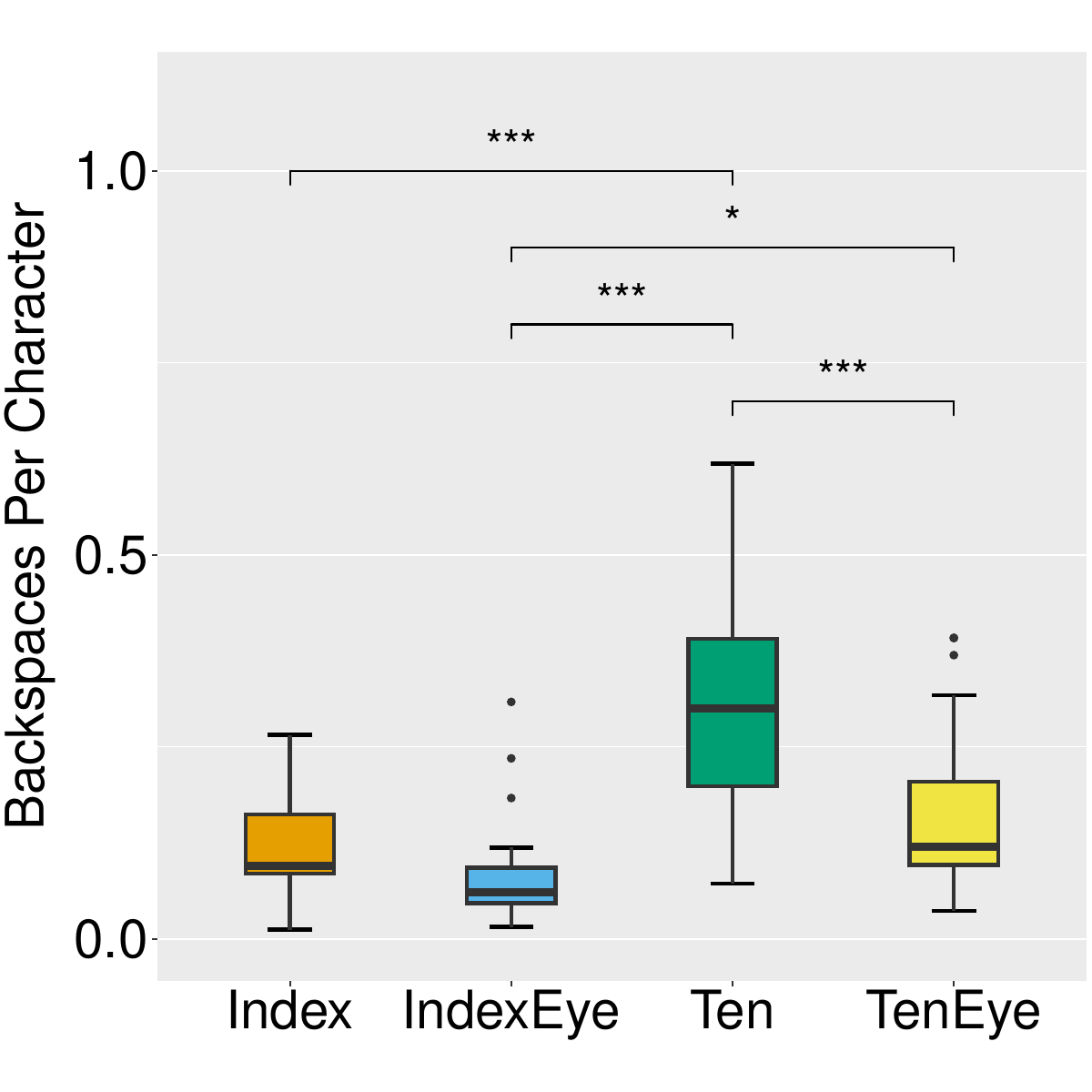}
\caption{Experiment 2: Entry rate (left), error rate (center), and backspaces per final output character (right). Brackets denote significant differences at $p<0.001$ (***), $p<0.01$ (**), and $p<0.05$ (*).\label{fig_exp2_wpm_cer}}
\Description{Three box plots showing the entry rate, character error rate, and backspaces per character in Experiment 2. Each plot shows the results for each of the four conditions.}
\end{figure}

We measured entry rate as in Experiment 1. Entry rates were 10.9 in \textsc{Index}, 9.4 in \textsc{IndexEye}, 7.1 in \textsc{Ten}, and 7.4 WPM in \textsc{TenEye} (Figure \ref{fig_exp2_wpm_cer} left). There were significant main effects for both the use of eye-tracking and the number of fingers used (Table \ref{table_stats_study2} left). As there was a significant interaction effect between the independent variables, we conducted post-hoc pairwise comparisons between conditions. All were significant except between \textsc{Ten} and \textsc{TenEye}.

\begin{figure}[tb]
\centering
\includegraphics[width=5cm]{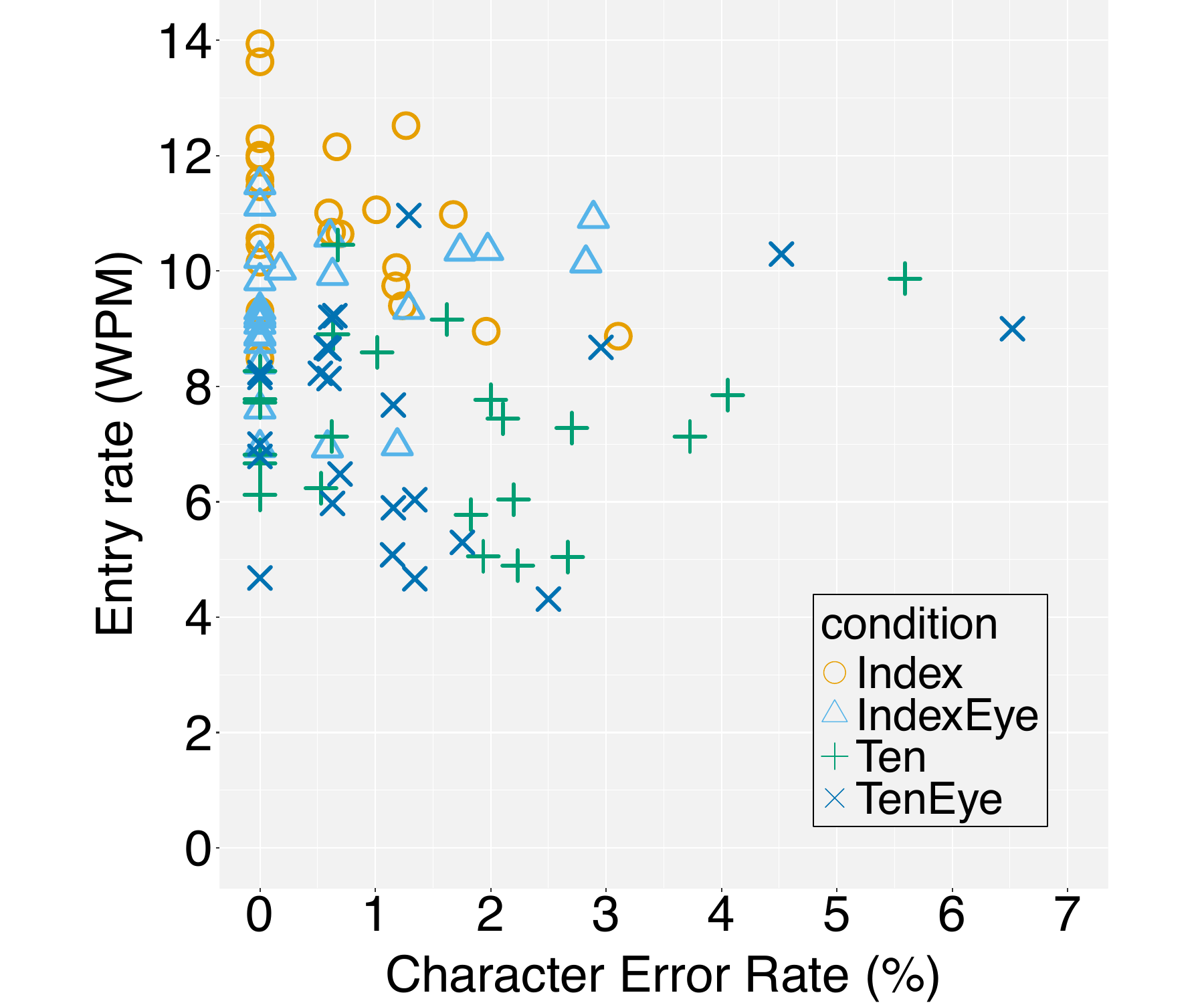}
\caption{Entry and error rate of each participant by condition in Experiment 2.\label{fig_exp2_scatter}}
\Description{Scatter plot showing how each participant did in each condition in Experiment 2 with respect to the character error rate (x-axis) and entry rate (y-axis).}
\end{figure}

\begin{table*}[tb]
\caption{Results from Experiment 2 (top), statistical tests for our two independent variables Eye and Fingers (middle), and post-hoc tests for dependent measures with significant interaction effects (bottom). Results are formatted as: mean $\pm$ sd [min, max]\label{table_stats_study2}.}
\begin{center}\small
\begin{tabular}{l l l l l }
\toprule
                          & & Entry rate (WPM) &  Error rate  (CER \%) & Backspaces per final character  \\
\midrule
Condition & \textsc{Index}           & 10.9 $\pm$ 1.4 [8.5, 13.9]                                     & 0.63 $\pm$ 0.82 [0.00, 3.11]                                & 0.12 $\pm$ 0.06 [0.01, 0.27]                             \\
&\textsc{IndexEye}        & \hspace{0.6mm} 9.4 $\pm$ 1.3 [6.9, 11.5]                                      & 0.58 $\pm$ 0.92 [0.00, 2.89]                                & 0.08 $\pm$ 0.07 [0.02, 0.31]                             \\
&\textsc{Ten}             & \hspace{0.6mm} 7.1 $\pm$ 1.7 [3.1, 10.5]                                      & 1.91 $\pm$ 2.21 [0.00, 9.63]                                & 0.30 $\pm$ 0.14 [0.07, 0.62]                             \\
&\textsc{TenEye}          & \hspace{0.6mm} 7.4 $\pm$ 1.9 [4.3, 11.0]                                      & 1.27 $\pm$ 1.54 [0.00, 6.52]                                & 0.16 $\pm$ 0.11 [0.04, 0.39]                             \\
\midrule
Independent & Eye                    & $F_{1,23}$\,=\,$5.8$,  $\eta_{G}^{2}$\,=\,$.04$, $p$\,<\,$.05$   & $F_{1,23}$\,=\,$1.1$,  $\eta_{G}^{2}$\,=\,$.01$, $p$\,=\,$.30$  & $F_{1,23}$\,=\,$23.6$, $\eta_{G}^{2}$\,=\,$.16$, $p$\,<\,$.0001$ \\
variable    & Fingers                & $F_{1,23}$\,=\,$78.9$, $\eta_{G}^{2}$\,=\,$.46$, $p$\,<\,$.0001$ & $F_{1,23}$\,=\,$15.0$, $\eta_{G}^{2}$\,=\,$.10$, $p$\,<\,$.001$ & $F_{1,23}$\,=\,$45.4$, $\eta_{G}^{2}$\,=\,$.32$, $p$\,<\,$.0001$ \\
            & Eye\,$\times$\,Fingers & $F_{1,23}$\,=\,$15.4$, $\eta_{G}^{2}$\,=\,$.08$, $p$\,<\,$.001$  & $F_{1,23}$\,=\,$0.7$,  $\eta_{G}^{2}$\,=\,$.01$, $p$\,=\,$.40$  & $F_{1,23}$\,=\,$9.3$,  $\eta_{G}^{2}$\,=\,$.07$, $p$\,<\,$.01$ \\
\midrule
Pairwise &             & \textsc{Index} $>$ \textsc{IndexEye}, $p$\,<\,$.01$     & n/a & \textsc{Index}    $\approx$ \textsc{IndexEye}, $p$\,=\,$.429$ \\
post-hoc &             & \textsc{Index} $>$ \textsc{Ten}, $p$\,<\,$.0001$        &     & \textsc{Index}    $<$       \textsc{Ten}, $p$\,<\,$.0001$     \\
tests    &             & \textsc{Index} $>$ \textsc{TenEye}, $p$\,<\,$.0001$     &     & \textsc{Index}    $\approx$ \textsc{TenEye}, $p$\,=\,$.247$   \\
         &             & \textsc{IndexEye} $>$ \textsc{Ten}, $p$\,<\,$.001$      &     & \textsc{IndexEye} $<$       \textsc{Ten}, $p$\,<\,$.0001$     \\
         &             & \textsc{IndexEye} $>$ \textsc{TenEye}, $p$\,<\,$.001$   &     & \textsc{IndexEye} $<$       \textsc{TenEye}, $p$\,<\,$.05$    \\
         &             & \textsc{Ten} $\approx$ \textsc{TenEye}, $p$\,=\,$1.00$  &     & \textsc{TenEye}   $<$       \textsc{Ten}, $p$\,<\,$.001$      \\
\bottomrule
\end{tabular}
\end{center}
\end{table*}

\subsubsection{Error rate.} As in Experiment 1, we measured typing errors using character error rate (CER). The CER was 0.63\% for \textsc{Index}, 0.58\% for \textsc{IndexEye}, 1.91\% for \textsc{Ten}, and 1.27\% for \textsc{TenEye} (Figure \ref{fig_exp2_wpm_cer} center). There was a significant main effect on the number of fingers used but not for the use of eye-tracking (Table \ref{table_stats_study2} center). Thus it seems that regardless of the use of eye-tracking filtering, errors increased for ten finger typing.

As shown in Figure \ref{fig_exp2_scatter}, numerous participants in \textsc{Ten} and \text{TenEye} had low entry rates of 4--8 WPM. Additionally, all instances of an error rate of 4\% or above were participants in these two conditions. While entry rate was overall slow in the ten-finger conditions, the addition of eye-tracking did seem to shift performance towards a lower error rate.

\subsubsection{Correction rate.} To measure error corrections, we calculated the number of backspaces per final output character for each phrase. The number of backspaces pressed per output character was 0.12 for \textsc{Index}, 0.08 for \textsc{IndexEye}, 0.30 for \textsc{Ten}, and 0.16 for \textsc{TenEye} (Figure \ref{fig_exp2_wpm_cer} right). There were significant main effects for both the use of eye-tracking and the number of fingers used (Table \ref{table_stats_study2} right). As there was a significant interaction effect, we conducted post-hoc pairwise comparisons between conditions. All were significant with the exception of between \textsc{Index} and \textsc{IndexEye}, and between \textsc{Index} and \textsc{TenEye}. Notably, eye-tracking did significantly reduce the number of corrections required in ten-finger typing. This suggests that our filtering feature was successful at preventing some erroneous keystrokes.

\subsubsection{Where participants looked.} We measured the percentage of the time users were looking at a key in the \textsc{Index} and \textsc{Ten} conditions such that the key would have been enabled in \textsc{IndexEye} and \textsc{TenEye}. This occurred for 83.8\% of key presses in \textsc{Index} and 84.6\% in \textsc{Ten}. Our implementation of eye-tracking eliminated 10.4\% of total key presses from \textsc{IndexEye} and 36.7\% from the \textsc{TenEye}. This indicates that user's gaze and typing style was relatively consistent between the \textsc{Index} and \textsc{IndexEye} conditions, since 83.8\% of key presses being valid in \textsc{Index} is similar to 89.6\% being valid in \textsc{IndexEye}. There was a much larger discrepancy between the \textsc{Ten} and \textsc{TenEye} conditions, with only 63.3\% of key presses being valid in \textsc{TenEye} as opposed to 84.6\% in  \textsc{Ten}. This may indicate that users were less careful with their hands in the \textsc{TenEye} condition and relied on the eye-tracking feature to filter out any extraneous key presses.

\begin{figure}[tb]
\includegraphics[width=14cm]{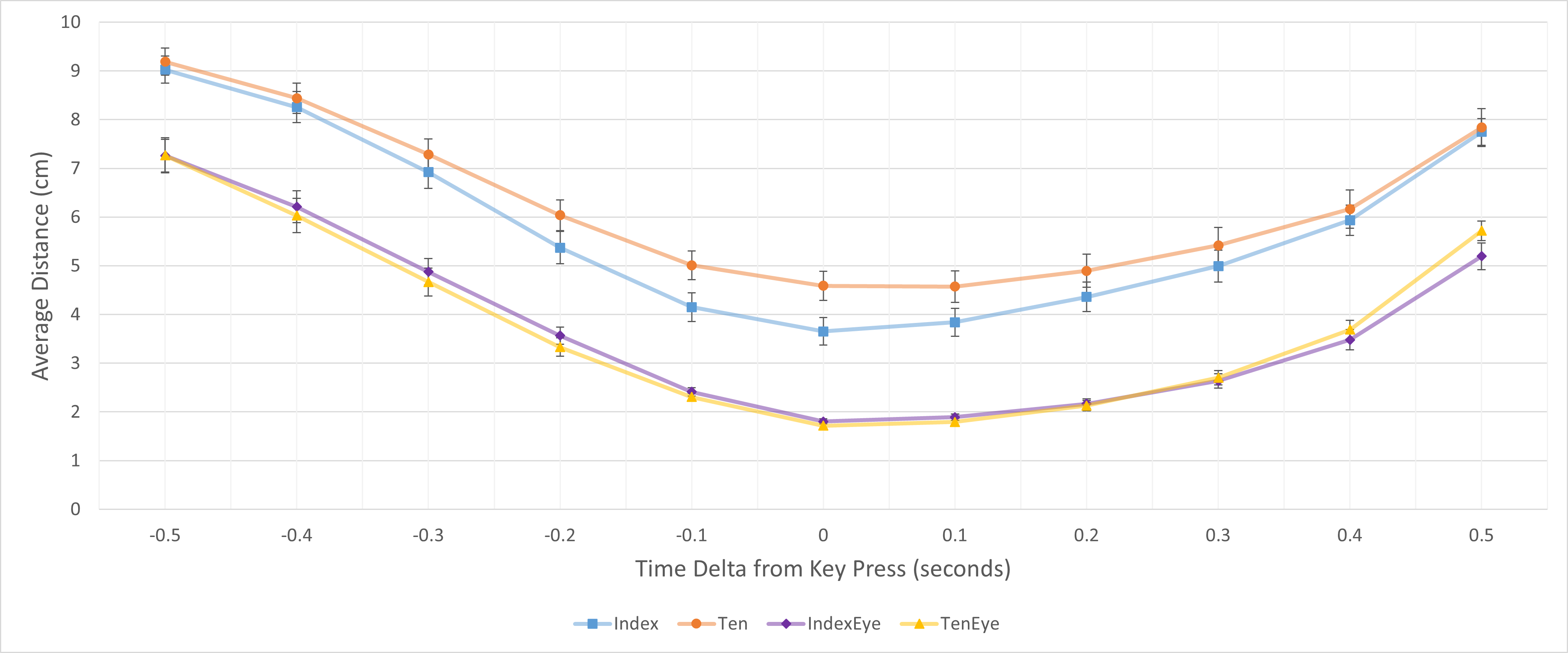}
\caption{The average distance in centimeters between  participants' eye gaze and the center of the pressed key over a time range before and after the key press. Error bars represent standard error.}
\label{eye_gaze_distance_exp2}
\Description{A line graph showing the average eye gaze distance of participants from a pressed key over a time offset from the key press.}
\end{figure}

Figure \ref{eye_gaze_distance_exp2} shows the average distance between the pressed key and eye gaze, before and after the key was pressed. In all conditions, the user looked most closely at the key at the instant it was pressed. Despite users tending to look more closely at the pressed key in the \textsc{IndexEye} and \textsc{TenEye} conditions, likely due to the restraints imposed by our eye-tracking feature, all four conditions followed the same pattern of eye movement. This supports our theory that users will naturally look where they intend to type. This indicates potential for incorporating this implicit gaze signal to improve a keyboard's interface or algorithms (e.g.~auto-correct or word predictions).

\subsubsection{Finger usage.} We analyzed how often participants used each finger to type to measure how well participants followed our instructions to type using both hands. In the index finger conditions, participants generally used both index fingers and preferred using their right index finger. In the \textsc{Index} condition, on average, participants used their right index finger 62\% of the time and ranged between 47\% and 75\% across participants. The results were similar in the \textsc{IndexEye} condition except that one participant used their right index finger exclusively even though we instructed all participants to try to use both hands. When we exclude that participant and average the others, we found that, similar to the \textsc{Index} condition, the right index finger was used 62\% of the time and ranged from 53\% to 78\% across participants.

\begin{figure}[tb]
\centering
\includegraphics[width=.48\columnwidth]{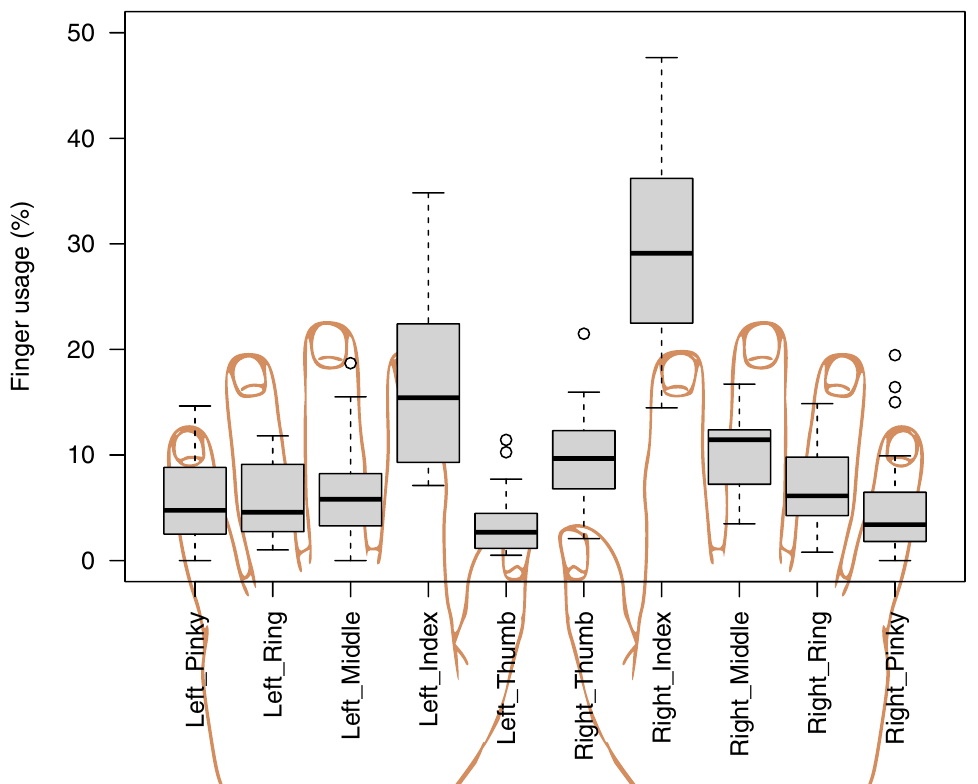}
\hspace{.2cm}
\includegraphics[width=.48\columnwidth]{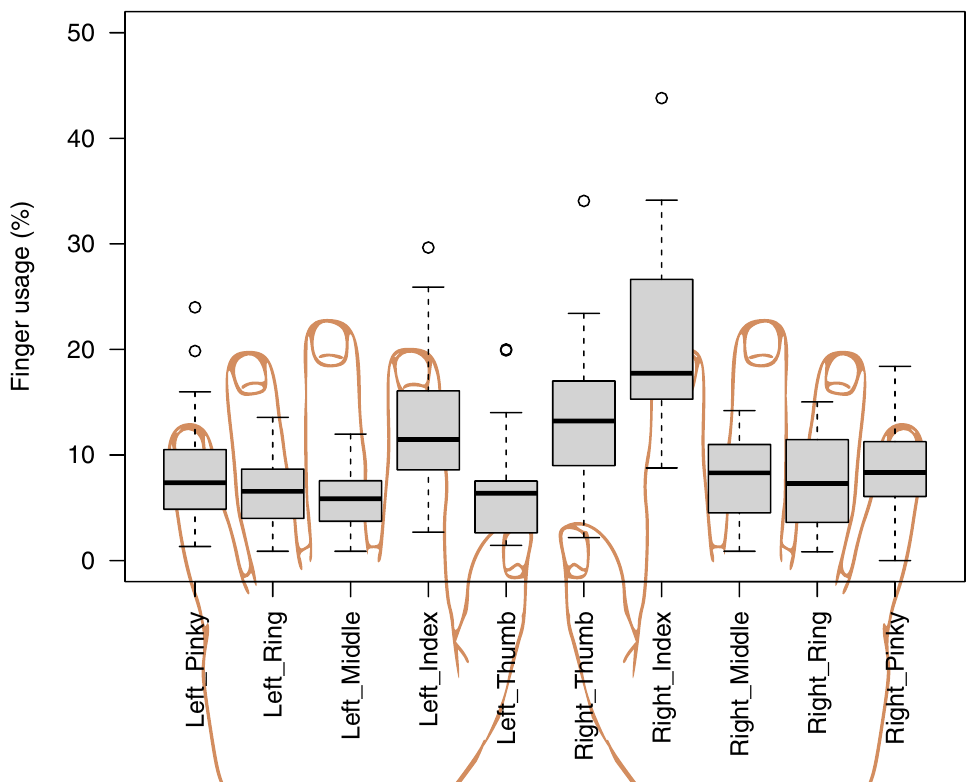}
\caption{Percentage of taps performed with each finger by participants in the \textsc{Ten} (left) and \textsc{TenEye} (right) conditions in Experiment 2.}
\label{fig_fingers_ten_exp2}
\Description{Boxplots showing which percentage taps were performed with each finger.}
\end{figure}

Figure \ref{fig_fingers_ten_exp2} shows which fingers participants used in the ten-finger conditions. Participants used all ten fingers at least once in these conditions with only a few exceptions (four participants in \textsc{Ten} and one in \textsc{TenEye}). Typing with index fingers was preferred with the right index finger being used the most frequently. In the \textsc{Ten} condition, on average, participants typed with their left and right index fingers 17\% and 30\% of the time respectively. Interestingly, in the \textsc{TenEye} condition, this tendency was less pronounced at 12\% and 21\% for the left and right index fingers respectively. This shows that the eye-tracking feature may have increased participants' confidence in typing with other fingers. 

\subsection{Subjective feedback}

\subsubsection{Likert statements.} 
\begin{figure}[tb]
\includegraphics[width=15cm]{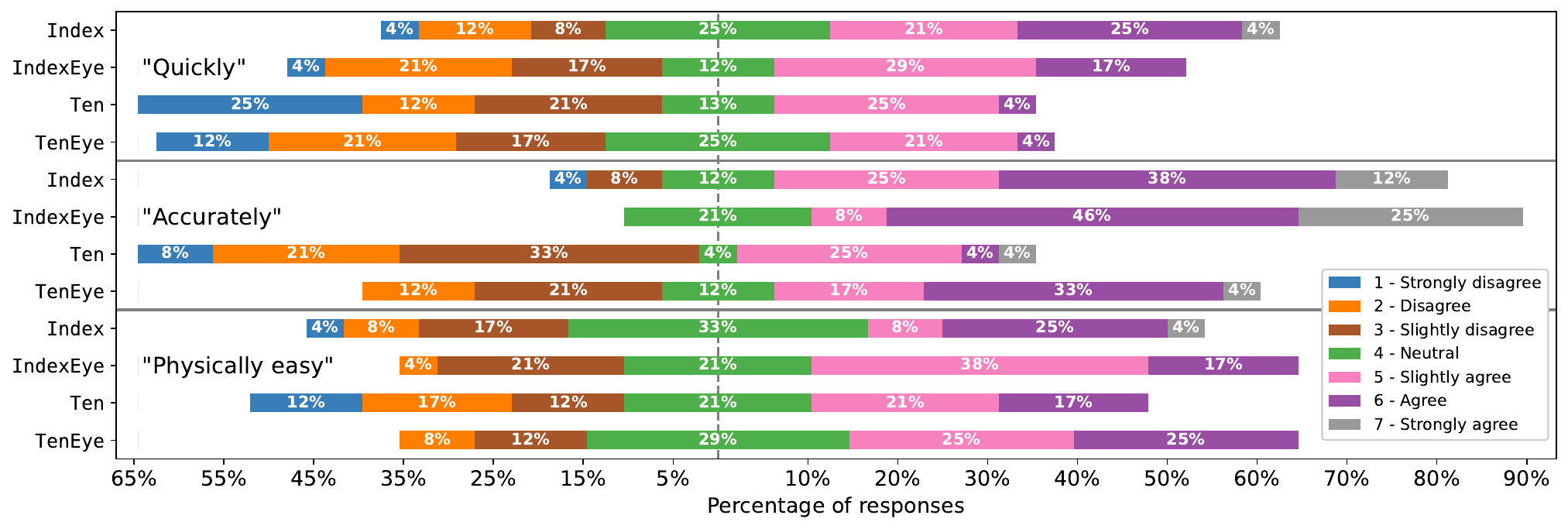}
\caption{Participants' Likert responses in Experiment 2 to the statements ``I entered text quickly'' (top), ``I entered text accurately'' (middle), and ``Entering text was physically easy and not tiring'' (bottom).}
\label{fig_exp2_likert}
\Description{Image showing the Likert ratings for three statements in Experiment 2.}
\end{figure}

After each condition, participants rated three statements on a 7-point Likert scale (Figure \ref{fig_exp2_likert}).
Participants' average ratings for ``I entered text quickly'' were: \textsc{Index} 4.4, \textsc{IndexEye} 3.9, \textsc{Ten} 3.1, and \textsc{TenEye} 3.3. The difference with respect to the number of fingers used was significant~($F_{1,23}$\,=\,$9.8, p$\,<\,$.01$). The difference with respect to the use of eye-tracking was not significant~($F_{1,23}$\,=\,$0.2, p$\,=\,$.68$). There was no significant interaction effect~($F_{1,23}$\,=\,$1.2, p$\,=\,$.27$). 

Participants' average ratings for ``I entered text accurately'' were:  \textsc{Index} 5.2, \textsc{IndexEye} 5.8, \textsc{Ten} 3.5, and \textsc{TenEye} 4.5. The differences were significant both with respect to the number of fingers ($F_{1,23}=41.5, p<.001$) and the use of eye-tracking ($F_{1,23}=8.5, p<.01$). There was no significant interaction effect ($F_{1,23}=2.1, p=.16$).   

Participants' average ratings for ``Entering text was physically easy and not tiring'' were: \textsc{Index} 4.2, \textsc{IndexEye} 4.4, \textsc{Ten} 3.7, and \textsc{TenEye} 4.5. There was no significant difference for the number of fingers ($F_{1,23}=0.6, p=.44$), the use of eye-tracking ($F_{1,23}=2.5, p=.13$), or the interaction between them ($F_{1,23}=2.4, p=.13$). 

\subsubsection{Condition ranking.} We asked participants to rank the conditions and to explain their reasoning. The \textsc{Index} condition was most preferred with 16 of 24 participants ranking it first. Some representative comments were: ``I didn't have to worry about other fingers getting in the way'', ``index fingers are both faster and more accurate for this'', and ``I prefer the comfort of being able to relax my other fingers''. The \textsc{Ten} condition was never ranked as the first choice and the last choice by 15 participants. 

\subsubsection{Likes and dislikes.} We coded participants' likes and dislikes as in Experiment 1.
Compared to the results of the condition ranking, these responses revealed a more nuanced result. 10 participants said they liked that \textsc{Index} only allowed two fingers but 7 disliked that limitation. After the \textsc{Ten} condition, 6 people said they liked being able to use all 10 fingers and 4 liked being able to use more fingers (but not all 10 fingers). Some representative comments were ``[index fingers] were mentally easier because there was less error with less fingers able to hit random buttons" and "the speed of typing with unrestricted 10 fingers was fun". People generally disliked how hard it was to type in the \textsc{Ten} condition with 15 people reporting unexpected keypresses and 12 reporting poor tracking particularly for certain fingers. For example, one participant disliked ``my fingers getting in the way while typing".

Eye tracking appeared to make ten finger typing less error-prone. After \textsc{TenEye}, 11 participants liked how eye tracking helped and only one disliked the unexpected keypresses in the condition. Despite the great reduction in unexpected keypresses in \textsc{TenEye}, there were two dislikes of tracking in general and nine dislikes of tracking for certain fingers. For example, one participant said ``middle and ring fingers don't track properly".  Only one participant ranked \textsc{TenEye} as their top ranked choice but 17 ranked it above \textsc{Ten}. \textsc{IndexEye} was ranked as the first choice by seven participants (the second most popular choice after \textsc{Index}) and was ranked above \textsc{IndexEye} by 17 participants. After the \textsc{IndexEye} condition, four people disliked that they couldn't use more fingers. The \textsc{IndexEye} and \textsc{TenEye} conditions each had four people who disliked being required to look at the spacebar. This suggests a possible interface improvement in which spacebar is excluded from the eye-tracking filtering.

\subsubsection{AR improvements.} Finally, we asked participants ``what needs to be improved or changed about the AR headset's hardware or software before you would be willing to use it in place of your mobile phone?'' Requested changes were: 
\begin{itemize}
    \item Improved hand-tracking accuracy and responsiveness, particularly for the ring finger (13 participants).
    \item Improvements related to the visual capabilities of the headset, specifically field of view (6 participants).
    \item Reduced eye-tracking latency (5 participants).
    \item Auto-correct or word predictions (4 participants).
    \item Increasing the keyboard sensitivity (1 participant).
    \item Reducing the bulk of the headset (1 participant).
\end{itemize}

\subsection{Discussion}

Our goals in Experiment 2 were to 1) examine if users typed faster in midair using all ten fingers, and 2) to see if eye-tracking could reduce accidental key presses. We found users had a significantly higher entry rate in the \textsc{Index} and \textsc{IndexEye} conditions than in the \textsc{Ten} and \textsc{TenEye} conditions. Users also expressed an overwhelming preference for the index finger conditions. Users performed poorly in the \textsc{Ten} condition as evidenced by lower entry rates compared to the index finger conditions and more backspaces per output character then all other conditions. Participants frequently complained in open feedback about the hand-tracking. They also reported difficulty pressing keys using their ring fingers and frustration with a lack of responsiveness of the hand-tracking.

Our hypothesis that users looked at the key they were attempting to press was supported by our eye-tracking data. We examined this by noting how often people were looking near the key they pressed in the \textsc{Index} and \textsc{Ten} conditions (conditions that did not use eye-tracking to disable keys). If we had disabled far away keys far as in \textsc{IndexEye} and \textsc{TenEye}, key presses would have been excluded 32.9\% and 36.0\% of the time in the \textsc{Index} and \textsc{Ten} conditions respectively. This suggests participants often looked where they were typing even when they were not required to do so. However, the \textsc{IndexEye} condition only excluded 10.4\% of key presses. This notably lower amount may be simply because participants were more careful to match their gaze to the key they were pressing, lowering the number of unintentional key presses. However, in the \textsc{TenEye} condition, 36.7\% of key presses occurred on disabled keys. This result is similar to the \textsc{Ten} condition and may show eye-tracking successfully filtered out unintentional key presses. However, it is important to note we do not know how many discarded key presses were due to unintentional presses being helpfully discarded versus erroneously discarded.

A common complaint during the eye-tracking conditions was having to look at the spacebar. Since the spacebar was large and thumbs often do not move far from it when typing, participants felt this unnecessarily slowed typing. In the future, keeping the spacebar active regardless of gaze position may improve results. Another common complaint was that the keys were unresponsive or difficult to trigger. We are uncertain if this was due to hand-tracking issues or the particular threshold distance we used for triggering keys. Reducing this distance would have made it easier to trigger keys, potentially improving performance. However, a shorter distance could also increase accidental key presses. The half-second cooldown may have also limited text entry speed in some circumstance, but given all participants in our studies typed substantially slower than 24\,WPM on average (i.e.~two characters per second), we think this did not unduly slow typing. Additional work could investigate if different key actuation details could help improve performance. 

There is also other evidence that eye-tracking was successful. Firstly, the addition of eye-tracking in \textsc{TenEye} significantly reduced the amount of backspaces per character (Figure \ref{fig_exp2_wpm_cer} right). Secondly, participants rated the stated ``I entered text accurately'' significantly higher in the eye-tracking conditions. Overall, our results demonstrate that eye-tracking is an effective tool that should be considered when implementing ten-finger midair typing systems.

\section{Limitations and Future Work}
Our entry rates in both experiments were undoubtedly limited by the deterministic nature of our keyboard. It is likely users could have typed substantially faster if they only had to aim approximately within a key or so of their desired letters. Powerful auto-correct might even be able to sift through the many false key triggers typical of ten-finger midair typing. However, even in the case a MR keyboard supporting fast typing via auto-correct, it would still need a fallback mode to support users entering exactly what they want. This can be necessary for certain types of content, or to avoid an ``auto-correct trap'' in which a user's desired and correct text is repeatedly auto ``corrected'' into incorrect text. Our work can help guide how an auto-correcting MR keyboard can best support fallback to deterministic input.

Past work on midair keyboards have commonly oriented keyboards either completely vertically   (e.g.~\cite{adhikary_midair,markussen2013selection}) or tilted slightly away from vertical (e.g.~\cite{adhikary2021text,dudley_augmented}). However, in most past studies, participants were standing and not seated as in our study. We opted instead for a 15 degree tilt from horizontal for both our studies. This felt more comfortable in pilot testing and more closely matches users' real-world typing experience on desktop keyboards which often have an upward tilt. A similar tilt of 20 degrees \cite{dudley_email2025} was used in another study which explored seated typing on or near a surface \cite{dudley2019strategies}. While this tilt appeared to work well for our participants, further exploration of user performance and preference with regards to keyboard tilt could be useful. 

Both experiments consisted of a single one-hour session. The majority of our participants had no prior mixed reality experience. For index-finger typing on a midair keyboard, it was reassuring to see that novices could achieve a functional level of text entry performance with little training. Furthermore, they did this on text requiring use of mixed case, punctuation, and numbers. However, it would be interesting to study virtual keyboard performance over a longer period of time to better understand whether better performance is attainable with practice. Long-term use would also allow exploring whether tuning a keyboard's models or thresholds can improve performance for a specific user.

Our participant population was young, tech savvy, and mostly male. While we suspect the differences we found between conditions were driven by challenges related to midair target selection and hand-tracking, we caution results could vary for other populations. 

An interesting avenue for future research would be to use auxiliary sensors (e.g.~the AR headset's microphone or a smartwatch) to detect taps on a physical surface instead of, or in addition to, finger collision with the keys. This could improve tap detection accuracy of virtual targets positioned on physical surfaces. In particular, the tactile feedback of actual contact with a surface could mitigate the problem of unintentional key presses in ten-finger typing.

While we a current commodity headset struggled to track all ten fingers, we found most participant complaints mostly revolved around certain fingers such as the ring and middle finger. An interesting future study might explore typing with a more limited set of fingers (e.g.~the index fingers plus thumbs and pinky fingers). This could allow users to use some fingers for common keys (e.g.~thumbs for spacebar and pinky finger for backspace).

In Experiment 2, we used gaze location to make a hard decision about which keys were active. This threshold was determined based on pilot testing and given the eye-tracking accuracy of our headset. It is possible a different threshold, or a more accurate eye-tracker, would improve performance of this feature. It would also be interesting to investigate improving location or tap detection accuracy via a softer, probabilistic fusion of the hand and eye-tracking signals. We have included in our supplementary materials\ifanon
~
\else
\footnote{\url{https://osf.io/q4n72}} 
\fi
detailed participant log files from both experiments. This includes when participants triggered individual keys as well as continuously sampled observations from the hand and eye tracker for Experiment 2. This data could allow exploring potential gains from incorporating such changes into future midair keyboards. We have also included the source code for the keyboards we used in both experiments.

\section{Conclusion}

Experiment 1 showed that a midair keyboard had significantly higher entry rates and was preferred over keyboards positioned on a surface. Entry rates were 12.2, 9.1, and 7.8 WPM in the \textsc{Midair}, \textsc{Table}, and \textsc{Wall} conditions respectively. Questionnaire results also showed that our midair keyboard, which was arranged such that participants could rest their arms on a table, was preferred and not particularly tiring compared to the keyboards positioned on a wall or table. Participant feedback was generally positive about the tactile feedback and ergonomics of typing on a virtual keyboard located on a table. However, at present, the hand-tracking of a current commodity headset (HoloLens 2) needs improvement to better support accurate deterministic interactions near surfaces.

In Experiment 2, we further explored midair typing comparing typing with ten fingers versus typing with index fingers. We tested using eye-tracking to help filter unintentional key presses. While we found eye-tracking reduced error corrections in ten-finger typing, it was still slower than index-finger typing. Entry rates were 10.9, 9.4, 7.1, and 7.4 WPM in the \textsc{Index}, \textsc{IndexEye}, \textsc{Ten}, and \textsc{TenEye} conditions respectively. In situations where midair ten-finger typing is desired, our results suggest eye-tracking filtering can reduce backspace use without negatively impacting users' entry or error rates. Subjective feedback also showed participants felt eye-tracking helped them type more accurately.

In summary, we found that using a commodity MR headset with no auxiliary devices or sensors, novices could type at a reasonable speed of around 11--12 WPM with a error rate below 1\%. They did this on realistic real-world text involving mixed case, punctuation, and numbers. Furthermore, they did this on a deterministic keyboard without auto-correct. At present, of the options we tested, typing using just the index fingers on a floating midair keyboard appears to be the most preferable option. 

% Note: these disappear if anonymous is set in documentclass
\begin{acks}
This material is based upon work supported by the NSF under Grant No. IIS-1909089. We thank Michigan Tech students in CS5641 from Spring 2024 for providing feedback on an initial draft of this paper.
\end{acks}

\bibliographystyle{ACM-Reference-Format}
\bibliography{references}
	
\end{document}